\documentclass[aps,twocolumn,showpacs]{revtex4}
\usepackage{graphicx}
\usepackage{dcolumn}
\usepackage{amsmath}
\usepackage{enumerate}
\begin{document}

\title{
Neutrinoless $\beta \beta $ decay transition matrix
elements within mechanisms involving light Majorana neutrinos, classical Majorons and
sterile neutrinos }

\author{P. K. Rath$^{1}$}
\author {R. Chandra$^{2}$}
\thanks {Corresponding author: ramesh\_dap@bbau.ac.in}
\author {K. Chaturvedi$^{3}$}
\author {P. Lohani$^{1}$}
\author {P. K. Raina$^{4}$}
\author {J. G. Hirsch$^{5}$}
\affiliation {
$^{1}$Department of Physics, University of Lucknow, Lucknow-226007, India}
\affiliation {
$^{2}$Department of Applied Physics, Babasaheb Bhimrao Ambedkar University,
Lucknow-226025, India}
\affiliation {
$^{3}$Department of Physics, Bundelkhand University, Jhansi-284128, India}
\affiliation {
$^{4}$Department of Physics, Indian Institute of Technology, Ropar,
Rupnagar - 140001, India}
\affiliation {
$^{5}$Instituto de Ciencias Nucleares, Universidad Nacional Aut\'{o}noma de
M\'{e}xico, 04510 M\'{e}xico, D.F., M\'{e}xico}

\date{\today}

\begin{abstract}
In the PHFB model, uncertainties in the nuclear transition matrix elements for 
the neutrinoless double-$\beta $ decay 
of $\ ^{94,96}$Zr, $^{98,100}$Mo, $^{104}$Ru, $^{110}$Pd, $^{128,130}$Te and 
$^{150}$Nd isotopes within mechanisms involving light 
Majorana neutrinos, classical Majorons and sterile neutrinos are statistically 
estimated by considering sets of sixteen (twenty-four) matrix elements calculated 
with four different parametrization of the pairing plus multipolar type of effective 
two-body interaction, two sets of form factors and two (three) different 
parameterizations of Jastrow type of short range correlations. In the mechanisms 
involving the light Majorana neutrinos and classical Majorons, the maximum uncertainty 
is about 15\% and in the scenario of sterile neutrinos, it varies in between 
approximately 4 (9)\%--20 (36)\% without(with) Jastrow short range correlations with 
Miller-Spencer parametrization, depending on the considered mass of the sterile neutrinos.
\end{abstract}


\maketitle

\section{INTRODUCTION}

In the last decade, the confirmation of neutrino flavor oscillations at
atmospheric, solar, reactor and accelerator neutrino sources \cite{garc08,bile12} 
and the reported observation of neutrinoless double beta 
$\left( \beta \beta \right) _{0\nu }$ decay 
\cite{klap01, klap04,feru02,aals02,zdes02,iann04} have together played
a great inspirational role in the advancement of a vast amount of
experimental as well as theoretical studies on nuclear double-$\beta $ decay
in general and $\left( \beta \beta \right) _{0\nu }$ decay in particular
\cite{avig08,verg12}. The former has provided information on the neutrino
mass square differences $\Delta $m$_{21}^{2}$ and $\Delta $m$_{31}^{2}$,
mixing angles $\theta _{12}$, $\theta _{23}$ and $\theta _{13}$ and possible
hierarchies in the neutrino mass spectrum \cite{fogl11}. In addition to hinting 
on the Majorana nature of neutrinos, the latter has also ascertained the role of
various mechanism in different gauge theoretical models \cite{klap06}. 
Presently, a number of projects for 
observing the $\left( \beta \beta \right)_{0\nu }$ decay of $^{48}$Ca (CANDLES), 
$^{76}$Ge (GERDA, MAJORANA), $^{82}$Se (SuperNEMO, Lucifer),$^{100}$Mo (MOON, Amore), 
$^{116}$Cd (COBRA), $^{130}$Te (CUORE), $^{136}$Xe (XMASS, EXO, KAMLand-Zen, 
NEXT), $^{150}$Nd (SNO++, SuperNEMO, DCBA/MTD)  \cite{bar10a,giul10,schw12} have been 
designed and hopefully, the reported observation of $\left( \beta \beta \right) _{0\nu }$
decay \cite{klap01, klap04} would be confirmed in the near future.

In the left-right symmetric model \cite{doi93,hilr96}, the three possible
mechanisms of $\left( \beta ^{-}\beta ^{-}\right) _{0\nu }$ decay are the
exchange of left handed light as well as heavy Majorana neutrinos and the exchange of 
right handed heavy Majorana neutrinos. Alternatively, the occurrence of lepton number 
violating Majoron accompanied $\left( \beta \beta \right) _{0\nu }$ decay is also a
possibility. Based on the most recent experimental evidences 
\cite{arno11,argy09,argy10,bara11}, regarding the observability of all the nine 
Majoron models  \cite{bame95,himj96}, it has been concluded that the study of classical 
Majoron models is the most preferred one.

In the short base line experiments \cite{agui01,agui10}, the indication of 
$\overline{\nu }_{\mu }\rightarrow \overline{\nu }_{e}$ conversion was explained with 
$0.2$ eV$<\Delta m^{2}<2$ eV and $10^{-3}<sin^{2}2\theta <4.10^{-2}$. New results of the reactor 
fluxes favor short base line oscillation \cite{ment11,muel11,hube11}. The confirmation of all 
these observations would imply the existence of more than three massive neutrinos \cite{guin11}. 
In Ref. \cite{bame95}, it was shown that the mixing of a light sterile neutrino (mass $\ll $ 1 eV) 
with a much heavier sterile neutrino (mass $\gg $ 1 GeV) would result in observable signals in 
current $\beta \beta $ decay experiments, as is the case in other interesting alternative 
scenarios \cite{pila93,bene05}. 

The study of $\left( \beta ^{-}\beta^{-}\right) _{0\nu }$ decay within mechanisms involving 
light Majorana neutrinos, classical Majorons and sterile neutrinos can be performed under a 
common theoretical formalism \cite{doi85,haxt84,tomo91}. In the mass mechanism, the 
contributions  of the pseudoscalar and weak magnetism terms of the recoil current can change 
the NTMEs $M^{\left( 0\nu \right) }$ up to 30\% in the QRPA \cite{simk99,verg02}, about 20\% in 
the interacting shell model (ISM) \cite{caur08} and 15\% in the interacting Boson model (IBM) 
\cite{bare09}. 

In the evaluation of NTMEs, the most desirable approach is to 
employ the successful large scale shell-model calculations \cite{caur08,stma,horo10,horo13}, 
if feasible. However, the QRPA \cite{voge86,civi87} and its extensions 
\cite{suho98,faes98} have emerged as the most employed models for explaining the observed 
suppression of $M_{2\nu }$ in addition to correlating the single-$\beta $ GT strengths 
and half-lives of ($\beta^{-}\beta ^{-}$)$_{2\nu }$ decay by including a large number 
of basis states in the model space. The necessity for the inclusion of nuclear
deformation has resulted in the employment of deformed QRPA \cite
{simk04,fang11,faes12}, projected-Hartree-Fock-Bogoliubov (PHFB) \cite
{chan09,rath10,rath12,rath13}, pseudo-SU(3) \cite{jghi95}, 
IBM \cite{bare09,iach11,bare13,yosi13} and energy density functional (EDF) 
\cite{rodr10} approaches in the calculation of NTMEs. 
Additionally,  there are many a possibilities for the inclusion of the model dependent 
form factors for the finite size of nucleons (FNS), short range correlations (SRC) 
\cite{mill76,wu85,jghi95,kor07a,kor07b,simk09}, and 
the value of axial vector current coupling constant $g_{A}$ \cite{mend11}. 
Each model has a different truncation scheme for the unmanageable Hilbert space, and employs
a variety of residual interactions, resulting in NTMEs $M^{(0\nu )}$, which are of the same
order of magnitude but not identical.

In the analysis of uncertainties in NTMEs for $\left( \beta ^{-}\beta
^{-}\right) _{0\nu }$ decay, the spread between the available calculated
results \cite{voge00} was translated in to an average of all the available
NTMEs, and the standard deviation was treated as the measure of the
theoretical uncertainty \cite{bahc04,avig05}. Bilenky and Grifols \cite
{bile02} have suggested that the possible observation of $\left( \beta \beta
\right) _{0\nu }$ decay in several nuclei could be employed to check the
calculated NTMEs in a model independent way by comparing the ratios of the
NTMEs-squared with the ratios of observed half-lives $T_{1/2}^{0\nu }$.
Model specific theoretical uncertainties have been analyzed in the QRPA approach
\cite{rodi03,suho05,rodi06}. Further, studies on uncertainties in NTMEs due to the 
SRC have also been preformed in Refs. \cite{kor07a,simk09,simk08}.

The main objective of the present work is to study the effects
of pseudoscalar and weak magnetism terms on the Fermi, Gamow-Teller (GT) and
tensorial NTMEs for the $\left( \beta ^{-}\beta ^{-}\right) _{0\nu }$ decay
of $^{94,96}$Zr, $^{98,100}$Mo, $^{104}$Ru, $^{110}$Pd, $^{128,130}$Te and $%
^{150}$Nd isotopes in the light Majorana neutrino mass mechanism. In addition, 
we investigate effects due to deformation, FNS and the SRC vis-a-vis the radial 
evolution of NTMEs. Uncertainties in NTMEs are calculated statistically by 
employing four different parametrizations of effective two-body interaction, 
form factors with two different parametrizations and three different parametrizations 
of the SRC. In the same theoretical formalism, 
the $\left( \beta ^{-}\beta ^{-}\right) _{0\nu }$ decay involving classical
Majorons and sterile neutrinos is also studied. The theoretical formalisms to
calculate the half-lives of the $\left( \beta ^{-}\beta ^{-}\right) _{0\nu }$
decay with induced currents \cite{simk99,verg02}, classical Majorons \cite
{doi85,simk99} and sterile neutrinos \cite{bene05} have already been reported.
Hence, we briefly outline the steps of the above derivations in Sec. II. In
Sec. III, we present the results and discuss them vis-a-vis the existing
calculations done in other nuclear models. Finally, the conclusions are
given in Sec. IV.

\section{THEORETICAL\ FRAMEWORK}

The detailed theoretical formalism required for the study of 
$\left( \beta ^{-}\beta^{-}\right) _{0\nu }$ decay due to the exchange of 
light Majorana neutrinos has been given by \v{S}imkovic\textit{\ et. al.} 
\cite{simk99} as well as Vergados \cite{verg02}. The observability of 
Majoron accompanied $\left( \beta ^{-}\beta^{-}\right) _{0\nu }$ decay 
in nine Majoron models \cite{bame95} has already been discussed by Hirsch 
\textit{et. al.} \cite{himj96}. Further, the 
$\left( \beta ^{-}\beta^{-}\right) _{0\nu }$ decay within the mechanism
involving sterile neutrinos has been given by Benes \textit{et al.} \cite
{bene05}. In the following, we present a brief out line of the required
theoretical formalism for the clarity in notations used in the present paper.

\subsection{Light Majoron neutrino mass mechanism}

In the Majorana neutrino mass mechanism, the half-life $T_{1/2}^{\left( 0\nu
\right) }$ for the 0$^{+}\to $0$^{+}$\ transition of $\left( \beta ^{-}\beta
^{-}\right) _{0\nu }$ decay due to the exchange of light Majorana neutrinos
between nucleons having finite size is given by \cite{simk99,verg02}

\begin{equation}
\left[ T_{1/2}^{\left( 0\nu \right) }(0^{+}\rightarrow 0^{+})\right]
^{-1}=G_{01}\left|\frac{\left\langle m_{\nu }\right\rangle }{m_{e}} 
M^{\left( 0\nu \right) }\right| ^{2},
\end{equation}
where
\begin{equation}
\left\langle m_{\nu }\right\rangle =\sum\nolimits_{i}^{\prime
}U_{ei}^{2}m_{i},\qquad \qquad m_{i}<10\text{ }eV,
\end{equation}

\begin{widetext}
\begin{equation}
G_{01}=\left[ \frac{2\left( G_{F}\text{ }g_{A}\right) ^{4}m_{e}^{9}}{64\pi
^{5}\text{ }\left( m_{e}R\right) ^{2}\ln \left( 2\right) }\right]
\int_{1}^{T+1}F_{0}\left( Z_{f},\varepsilon _{1}\right) F_{0}\left(
Z_{f},\varepsilon _{2}\right) p_{1}\text{ }p_{2}\text{ }\varepsilon
_{1}\varepsilon _{2}\text{ }d\varepsilon _{1},  \label{g01}
\end{equation}
\noindent and in the closure approximation, the NTME $M^{\left( 0\nu \right)
}$ is defined as
\begin{equation}
M^{\left( 0\nu \right) }=\sum_{n,m}\left\langle 0_{F}^{+}\left\| \left[ -%
\frac{H_{F}(r_{nm})}{g_{A}^{2}}+\mathbf{\sigma }_{n}\cdot \mathbf{\sigma }%
_{m}H_{GT}(r_{nm})+S_{nm}H_{T}(r_{nm})\right] \tau _{n}^{+}\tau
_{m}^{+}\right\| 0_{I}^{+}\right\rangle, \label{ntme}
\end{equation}
with
\begin{equation}
S_{nm}=3\left( \mathbf{\sigma }_{n}\cdot \widehat{\mathbf{r}}_{nm}\right)
\left( \mathbf{\sigma }_{m}\cdot \widehat{\mathbf{r}}_{nm}\right) -\mathbf{%
\sigma }_{n}\cdot \mathbf{\sigma }_{m}.
\end{equation}
\end{widetext}
The neutrino potentials associated with Fermi, Gamow-Teller (GT) and tensor
operators are given by
\begin{equation}
H_{\alpha }(r_{nm})=\frac{2R}{\pi }\int \frac{f_{\alpha }\left(
qr_{nm}\right) }{\left( q+\overline{A}\right) }\;h_{\alpha }(q)qdq,
\label{npot}
\end{equation}
where $f_{\alpha }\left( qr_{nm}\right) =j_{0}\left( qr_{nm}\right) $ and
$f_{\alpha }\left( qr_{nm}\right) =j_{2}\left(qr_{nm}\right) $  for $%
\alpha =$Fermi/$GT$ and tensor potentials, respectively. The effects due to 
the FNS are incorporated through the dipole form factors and the
form factor related functions $h_{F}(q)$, $h_{GT}(q)$ and $h_{T}(q)$ are written as
\begin{widetext}
\begin{eqnarray}
h_{F}(q) &=&g_{V}^{2}\left( q^{2}\right),   \label{eqhf} \\
h_{GT}(q) &=&\frac{g_{A}^{2}(q^{2})}{g_{A}^{2}}\left[ 1-\frac{2}{3}\frac{%
g_{P}(q^{2})q^{2}}{g_{A}(q^{2})2M_{p}}+\frac{1}{3}\frac{g_{P}^{2}(q^{2})q^{4}%
}{g_{A}^{2}(q^{2})4M_{P}^{2}}\right] +\frac{2}{3}\frac{g_{M}^{2}(q^{2})q^{2}%
}{g_{A}^{2}4M_{p}^{2}}  \nonumber \\
&\approx &\left( \frac{\Lambda _{A}^{2}}{q^{2}+\Lambda _{A}^{2}}\right)
^{4}\left[ 1-\frac{2}{3}\frac{q^{2}}{\left( q^{2}+m_{\pi }^{2}\right) }+%
\frac{1}{3}\frac{q^{4}}{\left( q^{2}+m_{\pi }^{2}\right) ^{2}}\right]
+\left( \frac{g_{V}}{g_{A}}\right) ^{2}\frac{\kappa ^{2}q^{2}}{6M_{p}^{2}}%
\left( \frac{\Lambda _{V}^{2}}{q^{2}+\Lambda _{V}^{2}}\right) ^{4},
\label{eqhgt} \\
h_{T}(q) &=&\frac{g_{A}^{2}(q^{2})}{g_{A}^{2}}\left[ \frac{2}{3}\frac{%
g_{P}(q^{2})q^{2}}{g_{A}(q^{2})2M_{p}}-\frac{1}{3}\frac{g_{P}^{2}(q^{2})q^{4}%
}{g_{A}^{2}(q^{2})4M_{P}^{2}}\right] +\frac{1}{3}\frac{g_{M}^{2}(q^{2})q^{2}%
}{g_{A}^{2}4M_{p}^{2}},  \nonumber \\
&\approx &\left( \frac{\Lambda _{A}^{2}}{q^{2}+\Lambda _{A}^{2}}\right)
^{4}\left[ \frac{2}{3}\frac{q^{2}}{\left( q^{2}+m_{\pi }^{2}\right) }-\frac{1%
}{3}\frac{q^{4}}{\left( q^{2}+m_{\pi }^{2}\right) ^{2}}\right] +\left( \frac{%
g_{V}}{g_{A}}\right) ^{2}\frac{\kappa ^{2}q^{2}}{12M_{p}^{2}}\left( \frac{%
\Lambda _{V}^{2}}{q^{2}+\Lambda _{V}^{2}}\right) ^{4},  \label{eqht}
\end{eqnarray}
\end{widetext}
where
\begin{eqnarray}
g_{V}(q^{2}) &=&g_{V}\left( \dfrac{\Lambda _{V}^{2}}{q^{2}+\Lambda _{V}^{2}}%
\right) ^{2}  \label{fm1v} \\
g_{A}(q^{2}) &=&g_{A}\left( \dfrac{\Lambda _{A}^{2}}{q^{2}+\Lambda _{A}^{2}}%
\right) ^{2}  \label{fm1a} 
\end{eqnarray}
\begin{eqnarray}
g_{P}(q^{2}) &=&\dfrac{2M_{p}g_{A}(q^{2})}{\left( q^{2}+m_{\pi }^{2}\right) }%
\left( \dfrac{\Lambda _{A}^{2}-m_{\pi }^{2}}{\Lambda _{A}^{2}}\right)
\label{fm1p} \\
g_{M}(q^{2}) &=&\kappa g_{V}\left( q^{2}\right)   \label{fm1m}
\end{eqnarray}
with $g_{V}=1.0$, $g_{A}=1.254$, $\kappa =\mu _{p}-\mu _{n}=3.70$, $\Lambda
_{V}=0.850$ GeV and $\Lambda _{A}=1.086$ GeV. The presence of
pseudoscalar and weak magnetism terms of the higher order currents (HOC) 
\cite{simk99}, results as seen by the consideration of Eqs. (\ref{ntme})--Eq. 
(\ref{fm1m}), in no change of the Fermi matrix element 
$M_{F}^{\left( 0\nu \right) }=-g_{A}^{2}M_{F-VV}^{\left( 0\nu \right) }$,
the addition of three new terms $M_{GT-AP}^{\left( 0\nu \right) }$, $%
M_{GT-PP}^{\left( 0\nu \right) }$, $M_{GT-MM}^{\left( 0\nu \right) }$ to
the conventional GT $M_{GT}^{\left( 0\nu \right) }=M_{GT-AA}^{\left( 0\nu
\right) }$ matrix element and the addition of three new terms 
$M_{T-AP}^{\left( 0\nu \right)}$, $M_{T-PP}^{\left( 0\nu \right) }$, 
$M_{T-MM}^{\left( 0\nu \right) }$ as the tensor matrix element 
$M_{T}^{\left( 0\nu \right) }$.

\begin{figure}[htbp]
\includegraphics [scale=0.68,angle=-90]{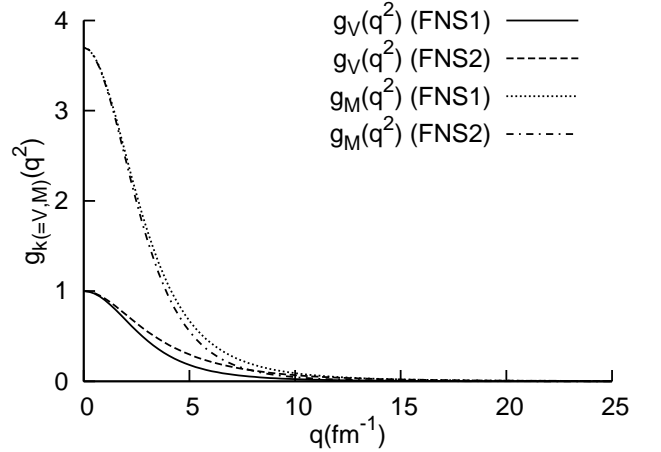}
\caption{Distribution of $g_{V}(q^{2})$ and $g_{M}(q^{2})$ for FNS1 and FNS2.}
\label{fig1}
\end{figure}
Consideration of internal structure of protons and neutrons suggests an
alternative parametrization of $g_{V}(q^{2})$ given by \cite{stol93}
\begin{equation}
g_{V}(q^{2})=F_{1}^{p}(q^{2})-F_{1}^{n}(q^{2})  \label{fm2v}
\end{equation}
where
\begin{eqnarray}
F_{1}^{p}(q^{2}) &=&\frac{1}{\left( 1+\frac{q2}{4M^{2}}\right) }\left(
\dfrac{\Lambda _{V}^{2}}{q^{2}+\Lambda _{V}^{2}}\right) ^{2}\left[ 1+\left(
1+\mu _{p}\right) \frac{q2}{4M^{2}}\right]  \\
F_{1}^{n}(q^{2}) &=&\frac{\mu _{n}}{\left( 1+\frac{q2}{4M^{2}}\right) }%
\left( \dfrac{\Lambda _{V}^{2}}{q^{2}+\Lambda _{V}^{2}}\right) ^{2}\left(
1-\xi _{n}\right) \frac{q2}{4M^{2}}
\end{eqnarray}
with $\mu _{p}=1.79$ nm, $\mu _{n}=-1.91$ nm and $\Lambda _{V}=0.84$ GeV.

In addition,
\begin{equation}
g_{M}(q^{2})=F_{2}^{p}(q^{2})-F_{2}^{n}(q^{2})   \label{fm2m}
\end{equation}
where
\begin{eqnarray}
F_{2}^{p}(q^{2}) &=&\frac{\mu _{p}}{\left( 1+\frac{q2}{4M^{2}}\right) }%
\left( \dfrac{\Lambda _{V}^{2}}{q^{2}+\Lambda _{V}^{2}}\right) ^{2} \\
F_{2}^{n}(q^{2}) &=&\frac{\mu _{n}}{\left( 1+\frac{q2}{4M^{2}}\right) }%
\left( \dfrac{\Lambda _{V}^{2}}{q^{2}+\Lambda _{V}^{2}}\right) ^{2}\left( 1+%
\frac{q2}{4M^{2}}\xi _{n}\right)
\end{eqnarray}
with
\begin{equation}
\xi _{n}=\frac{1}{1+\lambda _{n}\frac{q2}{4M^{2}}}
\end{equation}
and $\lambda _{n}=5.6$. We get two sets of form factors by considering Eq.
(\ref{fm1v})--Eq. (\ref{fm1m}) and Eqs. (\ref{fm1a},\ref{fm1p},\ref{fm2v},%
\ref{fm2m}), which are denoted by FNS1 and FNS2, respectively. In Fig. 1, we 
present the plot of $g_{V}(q^{2})$ and $g_{M}(q^{2})$ of FNS1 and FNS2, the 
shapes of which have definite relation with the magnitudes of NTMEs 
$M^{\left( 0\nu \right) }$. 

\subsection{Majoron accompanied $\left( \beta ^{-}\beta ^{-}\right) _{0\nu }$%
decay}

In the classical Majoron model, the inverse half-life $T_{1/2}^{\left( 0\nu
\phi \right) }$ for the 0$^{+}\rightarrow $0$^{+}$ transition of Majoron
emitting $\left( \beta ^{-}\beta ^{-}\phi \right) _{0\nu }$ decay is given
by \cite{doi85,simk99}
\begin{equation}
\lbrack T_{1/2}^{\left( 0\nu \phi \right) }\left( 0^{+}\rightarrow
0^{+}\right) ]^{-1}=\left| \left\langle g_{M}\right\rangle \right|
^{2} G_{0M} \left| M^{\left( 0\nu \phi \right) }\right| ^{2},
\end{equation}
where $\left\langle g_{M}\right\rangle $ is the effective Majoron--neutrino
coupling constant and the NTME $M^{\left( 0\nu \phi \right) }$ is same as
the $M^{\left( 0\nu \right) }$ for the exchange of light Majorana neutrinos.
The phase space factors $G_{0M}$ are evaluated by using

\begin{widetext}
\begin{equation}
G_{0M }=\left[ \frac{2\left( G_{F}\text{ }g_{A}\right) ^{4}m_{e}^{9}}{256\pi
^{7}\text{ }\left( m_{e}R\right) ^{2}\ln \left( 2\right) }\right]
\int_{1}^{T+1}F_{0}\left( Z_{f},\varepsilon _{1}\right) p_{1}\text{ }%
\varepsilon _{1}\text{ }d\varepsilon _{1}\int_{1}^{T+2-\varepsilon
_{1}}\left( T+2-\varepsilon _{1}-\varepsilon _{2}\right) F_{0}\left(
Z_{f},\varepsilon _{2}\right) p_{2}\text{ }\varepsilon _{2}d\varepsilon _{2},
\end{equation}
\end{widetext}
and have been calculated for all nuclei of general interest \cite{doi88,himj96}.

\subsection{Mechanism involving Sterile neutrinos}

The contribution of the sterile $\nu _{h}$ neutrino to the half-life 
$T_{1/2}^{\left( 0\nu \right) }$ for the $0^{+}\rightarrow
0^{+}$ transition of $\left( \beta ^{-}\beta ^{-}\right) _{0\nu }$ decay 
has been derived by considering the exchange of a Majorana neutrino between 
two nucleons and is given by \cite{bene05}
\begin{equation}
\lbrack T_{1/2}^{\left( 0\nu \right) }(0^{+}\rightarrow
0^{+})]^{-1}=G_{01}\left| U_{eh}^{2}\frac{m_{h}}{m_{e}}M^{0\nu
}(m_{h})\right| ^{2},
\end{equation}
where the phase space factor $G_{01}$ is the same as Eq. (\ref{g01}), $U_{eh}$
is the $\nu _{h}-\nu _{e}$ mixing matrix element and the NTME $M^{0\nu
}(m_{h})$ is written as
\begin{widetext}
\begin{equation}
M^{0\nu }(m_{h})=\left\langle 0_{F}^{+}\left\| \left[ -\frac{H_{F}\left(
m_{h,}r\right) }{g_{A}^{2}}+\mathbf{\sigma }_{n}\cdot \mathbf{\sigma }
_{m}H_{GT}\left( m_{h,}r\right) +S_{nm}H_{T}\left( m_{h,}r\right) \right]
\tau _{n}^{+}\tau _{m}^{+}\right\| 0_{I}^{+}\right\rangle   \label{eqst}
\end{equation}, 
\end{widetext}
In Eq. (\ref{eqst}), the neutrino potentials are of the form
\begin{equation}
H_{\alpha }\left( m_{h,}r\right) =\frac{2R}{\pi }\int_{0}^{\infty }\frac{%
f_{\alpha }(qr)h_{\alpha }(q^{2})q^{2}dq}{\sqrt{q^{2}+m_{h}^{2}}\left( \sqrt{%
q^{2}+m_{h}^{2}}+\overline{A}\right) },
\end{equation}
with the same $h_{\alpha }(q^{2})$ as given in Eqs. (\ref{eqhf})--Eq. 
(\ref{eqht}).

\subsection{Uncertainties in NTMEs within PHFB\ Model}

In the PHFB model, the calculation of the NTMEs $M_{k}^{\left( 0\nu \right) }
$ of the $\left( \beta ^{-}\beta ^{-}\right) _{0\nu }$ decay has already
been discussed in Ref. \cite{rath10}. Employing the HFB wave functions, one obtains the
following expression for the NTME $M_{k}^{\left( 0\nu \right) }$ of the $%
\left( \beta ^{-}\beta ^{-}\right) _{0\nu }$ decay corresponding to an
operator $O_{k}$ \cite{rath10}.
\begin{eqnarray}
M_{k}^{\left( 0\nu \right) } &=&\left[ n^{J_{f}=0}n^{Ji=0}\right] ^{-1/2}
\nonumber \\
&&\times \int\limits_{0}^{\pi }n_{(Z,N),(Z+2,N-2)}(\theta
)\sum\limits_{\alpha \beta \gamma \delta }\left\langle \alpha \beta \left|
O_{k}\right| \gamma \delta \right\rangle \nonumber \\
&&\times \sum\limits_{\varepsilon \eta }\frac{\left( f_{Z+2,N-2}^{(\pi
)*}\right) _{\varepsilon \beta }}{\left[ \left( 1+F_{Z,N}^{(\pi )}(\theta
)f_{Z+2,N-2}^{(\pi )*}\right) \right] _{\varepsilon \alpha }}  \nonumber \\
&&\times \frac{\left( F_{Z,N}^{(\nu )*}\right) _{\eta \delta }}{\left[
\left( 1+F_{Z.N}^{(\nu )}(\theta )f_{Z+2,N-2}^{(\nu )*}\right) \right]
_{\gamma \eta }}sin\theta d\theta.   \label{eqf}
\end{eqnarray}
The required amplitudes $(u_{im},v_{im})$ and expansion coefficients $%
C_{ij,m}$ of axially symmetric HFB intrinsic state ${|\Phi _{0}\rangle }$
with $K=0$ to evaluate the expressions $n^{J}$, $n_{(Z,N),(Z+2,N-2)}(\theta )
$, $f_{Z,N}$\ \ and $F_{Z,N}(\theta )$ \cite{rath10}, are obtained by
minimizing the expectation value of the effective Hamiltonian given by
\begin{equation}
H=H_{sp}+V(P)+V(QQ)+V(HH),
\end{equation}
in a basis consisting of a set of deformed states. The  details about 
the single particle Hamiltonian $H_{sp}$ as well as the pairing $V(P)$, 
quadrupole-quadrupole $V(QQ)$ and hexadecapole-hexadecapole  $V(HH)$ parts of 
the effective two-body interaction have been given in Ref. \cite{chan09}.
To perform a statistical analysis, sets of twenty-four NTMEs $M^{\left( 0\nu \right) }$ 
for $\left( \beta ^{-}\beta ^{-}\right) _{0\nu }$ decay are evaluated using 
Eq. (\ref{eqf}) in conjunction with four different parametrization of the two
body effective interaction, two sets of form factors and three different 
parametrizations of the SRC. The
details about the four different parametrizations have already been given in
Refs. \cite{chan09,rath10}. However, a brief discussion about them is
presented here for completeness shake.

The strengths of the proton-proton, the neutron-neutron and the
proton-neutron parts of the $V(QQ)$ are denoted by $\chi _{2pp},\chi _{2nn}$
and $\chi _{2pn}$, respectively. In Refs. \cite{chan05,sing07}, it has been
shown that the experimental excitation energy of the $\ $2$^{+}$ state, \ $%
E_{2^{+}}$ can be fitted by taking equal strengths of the like particle 
components of the QQ interaction i.e. $\chi _{2pp}=\chi_{2nn}=0.0105$ MeV 
\textit{b}$^{-4}$ and by varying the strength of the proton-neutron component 
of the $QQ$ interaction $\chi _{2pn}$. In Ref. \cite{chan09}, it was also feasible
to employ an alternative isoscalar parametrization of the
quadrupole-quadrupole interaction, by taking $\chi _{2pp}=\chi _{2nn}=\chi
_{2pn}/2$ and the three parameters were varied together to fit $E_{2^{+}}$.
These two alternative parameterizations of the quadrupole-quadrupole
interaction were referred to as \textit{PQQ}1 and \textit{PQQ}2. Two additional
parametrizations, namely \textit{PQQHH}1 and \textit{PQQHH}2 were obtained with the
inclusion of the hexadecapolar {\it HH} part of the effective interaction.
Presently, we consider a form of Jastrow short range correlations simulating
the effects of Argonne V18, CD-Bonn potentials in the self-consistent CCM
 \cite{simk09}, given by

\begin{equation}
f(r)=1-ce^{-ar^{2}}(1-br^{2}),
\end{equation}
where $a=1.1$ $fm^{-2}$, $1.59$ $fm^{-2}$, $1.52$ $fm^{-2}$, $b=0.68$ $%
fm^{-2}$, $1.45$ $fm^{-2}$, $1.88$ $fm^{-2}$ and $c=1.0$, $0.92$, $0.46$ for
Miller and Spencer parametrization, Argonne NN, CD-Bonn potentials, which
are denoted by SRC1, SRC2 and SRC3, respectively. Finally, the uncertainties
associated with the NTMEs $M^{(0\nu )}$ for $(\beta ^{-}\beta ^{-})_{0\nu }$
decay are evaluated by calculating the mean and standard deviation given by
\begin{equation}
\overline{M}^{(0\nu )}=\frac{\sum_{i=1}^{N}M_{i}^{(0\nu )}}{N},
\end{equation}
and
\begin{equation}
\Delta \overline{M}^{(0\nu )}=\frac{1}{\sqrt{N-1}}\left[
\sum_{i=1}^{N}\left( \overline{M}^{(0\nu )}-M_{i}^{(0\nu )}\right)
^{2}\right] ^{1/2}.
\end{equation}

\begin{table*}[htbp]
\caption{Decomposition of NTMEs $M^{(0\nu )}$ for the 
$\left( \beta ^{-}\beta ^{-}\right) _{0\nu }$ decay of $^{100}$Mo including higher order 
currents (HOC) with (a) FNS1, (b) FNS2 and SRC (HOC+SRC) for the {\it PQQ}1 parameterization.}
\label{tab1}
\begin{tabular}{lrrrrrrrrrrrrrrrrrrrrrr}
\hline\hline
NTMEs & ~~~~~~~ & FNS &~~~~~~  & HOC &~~~~~  &  &  &  & \multicolumn{5}{c}{HOC+SRC} &  &  &  &
~~~~~~~~& \multicolumn{5}{c}{HOC+SRC $(\overline{A}/2)$} \\ \cline{9-14}\cline{19-23}
&  &  &  &  &  &  &  &  & SRC1 &~~~~~  & SRC2 &~~~~~  & SRC3 &  &  &  &  & SRC1 &~~~~~  &
SRC2 &~~~~~  & SRC3 \\ \hline
$M_{F}^{(0\nu )}$ &  & \multicolumn{1}{c}{\small (a)} &  & {\small 2.1484} &
&  &  &  & {\small 1.8911} &  & {\small 2.1492} &  & {\small 2.2216} &  &  &
&  & {\small 2.0691} &  & {\small 2.3412} &  & {\small 2.4168} \\
&  & \multicolumn{1}{c}{\small (b)} &  & {\small 2.2034} &  &  &  &  &
{\small 1.9152} &  & {\small 2.1883} &  & {\small 2.2707} &  &  &  &  &
{\small 2.0943} &  & {\small 2.3817} &  & {\small 2.4673} \\
$M_{GT-AA}^{(0\nu )}$ &  & \multicolumn{1}{c}{} &  & {\small -6.3815} &  &
&  &  & {\small -5.4584} &  & {\small -6.3022} &  & {\small -6.5663} &  &  &
&  & {\small -5.9813} &  & {\small -6.8682} &  & {\small -7.1424} \\
$M_{GT-PP}^{(0\nu )}$ &  & \multicolumn{1}{c}{} &  & {\small -0.4503} &  &
&  &  & {\small -0.2962} &  & {\small -0.4054} &  & {\small -0.4510} &  &  &
&  & {\small -0.3060} &  & {\small -0.4177} &  & {\small -0.4640} \\
$M_{GT-AP}^{(0\nu )}$ &  & \multicolumn{1}{c}{} &  & {\small 1.5521} &  &  &
&  & {\small 1.1518} &  & {\small 1.4644} &  & {\small 1.5810} &  &  &  &  &
{\small 1.2013} &  & {\small 1.5222} &  & {\small 1.6413} \\
$M_{GT-MM}^{(0\nu )}$ &  & \multicolumn{1}{c}{\small (a)} &  & {\small %
-0.2370} &  &  &  &  & {\small -0.1192} &  & {\small -0.1832} &  & {\small %
-0.2201} &  &  &  &  & {\small -0.1239} &  & {\small -0.1892} &  & {\small %
-0.2266} \\
&  & \multicolumn{1}{c}{\small (b)} &  & {\small -0.2311} &  &  &  &  &
{\small -0.1222} &  & {\small -0.1850} &  & {\small -0.2187} &  &  &  &  &
{\small -0.1269} &  & {\small -0.1909} &  & {\small -0.2251} \\
$M_{GT}^{(0\nu )}$ &  & \multicolumn{1}{c}{\small (a)} &  & {\small -5.5167}
&  &  &  &  & {\small -4.7220} &  & {\small -5.4265} &  & {\small -5.6564} &
&  &  &  & {\small -5.2098} &  & {\small -5.9529} &  & {\small -6.1918} \\
&  & \multicolumn{1}{c}{\small (b)} &  & {\small -5.5107} &  &  &  &  &
{\small -4.7250} &  & {\small -5.4283} &  & {\small -5.6549} &  &  &  &  &
{\small -5.2128} &  & {\small -5.9546} &  & {\small -6.1902} \\
$M_{T-PP}^{(0\nu )}$ &  & \multicolumn{1}{c}{} &  & {\small -0.0227} &  &  &
&  & {\small -0.0230} &  & {\small -0.0235} &  & {\small -0.0234} &  &  &  &
& {\small -0.0236} &  & {\small -0.0241} &  & {\small -0.0240} \\
$M_{T-AP}^{(0\nu )}$ &  & \multicolumn{1}{c}{} &  & {\small 0.0692} &  &  &
&  & {\small 0.0700} &  & {\small 0.0710} &  & {\small 0.0709} &  &  &  &  &
{\small 0.0718} &  & {\small 0.0729} &  & {\small 0.0728} \\
$M_{T-MM}^{(0\nu )}$ &  & \multicolumn{1}{c}{\small (a)} &  & {\small 0.0059}
&  &  &  &  & {\small 0.0060} &  & {\small 0.0062} &  & {\small 0.0062} &  &
&  &  & {\small 0.0061} &  & {\small 0.0064} &  & {\small 0.0064} \\
&  & \multicolumn{1}{c}{\small (b)} &  & {\small 0.0058} &  &  &  &  &
{\small 0.0058} &  & {\small 0.0060} &  & {\small 0.0060} &  &  &  &  &
{\small 0.0059} &  & {\small 0.0062} &  & {\small 0.0062} \\
$M_{T}^{(0\nu )}$ &  & \multicolumn{1}{c}{\small (a)} &  & {\small 0.0524} &
&  &  &  & {\small 0.0529} &  & {\small 0.0538} &  & {\small 0.0537} &  &  &
&  & {\small 0.0543} &  & {\small 0.0552} &  & {\small 0.0551} \\
&  & \multicolumn{1}{c}{\small (b)} &  & {\small 0.0522} &  &  &  &  &
{\small 0.0528} &  & {\small 0.0536} &  & {\small 0.0535} &  &  &  &  &
{\small 0.0542} &  & {\small 0.0550} &  & {\small 0.0549} \\
$\left| M^{(0\nu )}\right| $ &  & \multicolumn{1}{c}{\small (a)} &  &
{\small 6.8305} &  &  &  &  & {\small 5.8716} &  & {\small 6.7394} &  &
{\small 7.0155} &  &  &  &  & {\small 6.4712} &  & {\small 7.3865} &  &
{\small 7.6736} \\
&  & \multicolumn{1}{c}{\small (b)} &  & {\small 6.8597} &  &  &  &  &
{\small 5.8902} &  & {\small 6.7664} &  & {\small 7.0454} &  &  &  &  &
{\small 6.4904} &  & {\small 7.4142} &  & {\small 7.7044} \\ \hline\hline
\end{tabular}
\end{table*}

\section{RESULTS\ AND\ DISCUSSIONS}

In the present work, we use the same wave functions as used in the earlier
works \cite{rath10,rath12,chan05,sing07}. It has been already shown that 
\cite{rath10}, the experimental excitation energies of $2^{+}$ state $%
E_{2^{+}}$ \cite{saka84} can be reproduced to about 98\% accuracy by
adjusting the proton-neutron quadrupolar correlation strength parameter $%
\chi _{2pn}$ or $\chi _{2pp}$. The maximum change in $E_{4^{+}}$ and $%
E_{6^{+}}$ energies with respect to \textit{PQQ}1 interaction \cite
{chan05,sing07} is about 8\% and 31\%, respectively. Further, the reduced $%
B(E2$:$0^{+}\to 2^{+})$ transition probabilities, deformation parameters $%
\beta _{2}$, static quadrupole moments $Q(2^{+})$ and gyromagnetic factors $%
g(2^{+})$ change by about 20\%, 10\% (except for $^{94}$Zr and \textit{PQQ}2
interaction), 27\% and 12 \% (except for $^{94,96}$Zr and \textit{PQQ}2
interaction), respectively, and are usually in an overall agreement with the
experimental data \cite{rama01,ragh89}. The experimental result for the
quadrupole moment $Q(2^{+})$ of $^{96}$Zr is not available and the
theoretical results exhibit a sign change for the \textit{PQQ}2 interaction.
The theoretically calculated NTMEs $M_{2\nu }$ for the $0^{+}\rightarrow
0^{+}$ transition also change up to approximately 11\% except $^{94}$Zr, $%
^{128,130}$Te, and $^{150}$Nd isotopes, for which the changes are
approximately 42$\%$, 21$\%,$ 24\% and 18$\%$ respectively. The change in
the ratio for deformation effect $D_{2\nu }$ is approximately 18\% but for $%
^{94}$Zr in which case the change is approximately 30\%.

\begin{table*}[htbp]
\caption{Calculated NTMEs $M^{(0\nu )}$ in the PHFB model with four
different parameterization of effective two-body interaction, namely 
(a) \textit{PQQ}1, (b) \textit{PQQHH}1, (c) \textit{PQQ}2 and (d) \textit{PQQHH}2, 
two different parametrizations of form factors and three different
parameterizations of Jastrow type of SRC for the $\left( \beta ^{-}\beta
^{-}\right) _{0\nu }$ decay of $^{94,96}$Zr, $^{98,100}$Mo, $^{104}$Ru, $^{110}$%
Pd, $^{128,130}$Te and $^{150}$Nd isotopes due to the exchange of light
Majorana neutrinos.} 
\label{tab2}
\begin{tabular}{cccccccccccccccccccccccccccc}
\hline\hline
{\small Nuclei} &~~~~~~  &  & {\small HOC1} &~~~  &  & \multicolumn{5}{c}{\small %
HOC1+SRC} & ~~~~ &~~~~  & \multicolumn{5}{c}{{\small HOC1+SRC(}$\overline{A}/2$%
{\small )}} &  &  & ~~~~~ & {\small HOC2} &~~~  & \multicolumn{5}{c}{\small HOC2+SRC}
\\ \cline{6-11}\cline{14-18}\cline{24-28}
&  &  &  &  &  & {\small SRC1} &  & {\small SRC2} &  & {\small SRC3} &  &  &
{\small SRC1} &  & {\small SRC2} &  & {\small SRC3} &  &  &  &  &  & {\small %
SRC1} &  & {\small SRC2} &  & {\small SRC3} \\ \hline
$^{94}${\small Zr} & {\small (a)} &  & {\small 4.3108} &  &  & {\small 3.6776%
} &  & {\small 4.2364} &  & {\small 4.4187} &  &  & {\small 4.0491} &  &
{\small 4.6377} &  & {\small 4.8272} &  &  &  & {\small 4.3293} &  & {\small %
3.6892} &  & {\small 4.2533} &  & {\small 4.4376} \\
& {\small (b)} &  & {\small 3.9577} &  &  & {\small 3.3702} &  & {\small %
3.8874} &  & {\small 4.0567} &  &  & {\small 3.7047} &  & {\small 4.2496} &
& {\small 4.4255} &  &  &  & {\small 3.9747} &  & {\small 3.3809} &  &
{\small 3.9030} &  & {\small 4.0741} \\
& {\small (c)} &  & {\small 4.1867} &  &  & {\small 3.6645} &  & {\small %
4.1359} &  & {\small 4.2867} &  &  & {\small 4.0619} &  & {\small 4.5584} &
& {\small 4.7151} &  &  &  & {\small 4.2020} &  & {\small 3.6741} &  &
{\small 4.1499} &  & {\small 4.3023} \\
& {\small (d)} &  & {\small 3.7598} &  &  & {\small 3.1949} &  & {\small %
3.6914} &  & {\small 3.8541} &  &  & {\small 3.5116} &  & {\small 4.0347} &
& {\small 4.2038} &  &  &  & {\small 3.7762} &  & {\small 3.2051} &  &
{\small 3.7064} &  & {\small 3.8708} \\
&  &  &  &  &  &  &  &  &  &  &  &  &  &  &  &  &  &  &  &  &  &  &  &  &  &
&  \\
$^{96}${\small Zr} & {\small (a)} &  & {\small 3.1099} &  &  & {\small 2.6244%
} &  & {\small 3.0489} &  & {\small 3.1881} &  &  & {\small 2.8994} &  &
{\small 3.3469} &  & {\small 3.4916} &  &  &  & {\small 3.1241} &  & {\small %
2.6332} &  & {\small 3.0619} &  & {\small 3.2026} \\
& {\small (b)} &  & {\small 3.0805} &  &  & {\small 2.5755} &  & {\small %
3.0155} &  & {\small 3.1602} &  &  & {\small 2.8321} &  & {\small 3.2959} &
& {\small 3.4464} &  &  &  & {\small 3.0950} &  & {\small 2.5845} &  &
{\small 3.0287} &  & {\small 3.1751} \\
& {\small (c)} &  & {\small 2.9718} &  &  & {\small 2.5049} &  & {\small %
2.9128} &  & {\small 3.0467} &  &  & {\small 2.7665} &  & {\small 3.1965} &
& {\small 3.3357} &  &  &  & {\small 2.9855} &  & {\small 2.5134} &  &
{\small 2.9253} &  & {\small 3.0606} \\
& {\small (d)} &  & {\small 2.8923} &  &  & {\small 2.4133} &  & {\small %
2.8301} &  & {\small 2.9673} &  &  & {\small 2.6533} &  & {\small 3.0926} &
& {\small 3.2353} &  &  &  & {\small 2.9060} &  & {\small 2.4219} &  &
{\small 2.8426} &  & {\small 2.9813} \\
&  &  &  &  &  &  &  &  &  &  &  &  &  &  &  &  &  &  &  &  &  &  &  &  &  &
&  \\
$^{98}${\small Mo} & {\small (a)} &  & {\small 7.0201} &  &  & {\small 6.0809%
} &  & {\small 6.9235} &  & {\small 7.1942} &  &  & {\small 6.7410} &  &
{\small 7.6293} &  & {\small 7.9107} &  &  &  & {\small 7.0483} &  & {\small %
6.0988} &  & {\small 6.9495} &  & {\small 7.2231} \\
& {\small (b)} &  & {\small 6.5208} &  &  & {\small 5.6151} &  & {\small %
6.4261} &  & {\small 6.6871} &  &  & {\small 6.1957} &  & {\small 7.0506} &
& {\small 7.3221} &  &  &  & {\small 6.5477} &  & {\small 5.6320} &  &
{\small 6.4507} &  & {\small 6.7146} \\
& {\small (c)} &  & {\small 7.0539} &  &  & {\small 6.1054} &  & {\small %
6.9563} &  & {\small 7.2296} &  &  & {\small 6.7656} &  & {\small 7.6626} &
& {\small 7.9468} &  &  &  & {\small 7.0824} &  & {\small 6.1234} &  &
{\small 6.9826} &  & {\small 7.2589} \\
& {\small (d)} &  & {\small 6.4528} &  &  & {\small 5.5523} &  & {\small %
6.3582} &  & {\small 6.6178} &  &  & {\small 6.1261} &  & {\small 6.9758} &
& {\small 7.2456} &  &  &  & {\small 6.4796} &  & {\small 5.5692} &  &
{\small 6.3828} &  & {\small 6.6451} \\
&  &  &  &  &  &  &  &  &  &  &  &  &  &  &  &  &  &  &  &  &  &  &  &  &  &
&  \\
$^{100}${\small Mo} & {\small (a)} &  & {\small 6.8305} &  &  & {\small %
5.8716} &  & {\small 6.7394} &  & {\small 7.0155} &  &  & {\small 6.4712} &
& {\small 7.3865} &  & {\small 7.6736} &  &  &  & {\small 6.8597} &  &
{\small 5.8902} &  & {\small 6.7664} &  & {\small 7.0454} \\
& {\small (b)} &  & {\small 6.5255} &  &  & {\small 5.5826} &  & {\small %
6.4329} &  & {\small 6.7043} &  &  & {\small 6.1305} &  & {\small 7.0272} &
& {\small 7.3096} &  &  &  & {\small 6.5538} &  & {\small 5.6005} &  &
{\small 6.4589} &  & {\small 6.7333} \\
& {\small (c)} &  & {\small 6.8843} &  &  & {\small 5.9193} &  & {\small %
6.7925} &  & {\small 7.0704} &  &  & {\small 6.5244} &  & {\small 7.4454} &
& {\small 7.7344} &  &  &  & {\small 6.9137} &  & {\small 5.9379} &  &
{\small 6.8196} &  & {\small 7.1005} \\
& {\small (d)} &  & {\small 5.8838} &  &  & {\small 5.0268} &  & {\small %
5.8000} &  & {\small 6.0467} &  &  & {\small 5.5173} &  & {\small 6.3327} &
& {\small 6.5892} &  &  &  & {\small 5.9095} &  & {\small 5.0431} &  &
{\small 5.8237} &  & {\small 6.0730} \\
&  &  &  &  &  &  &  &  &  &  &  &  &  &  &  &  &  &  &  &  &  &  &  &  &  &
&  \\
$^{104}${\small Ru} & {\small (a)} &  & {\small 4.9041} &  &  & {\small %
4.1810} &  & {\small 4.8420} &  & {\small 5.0500} &  &  & {\small 4.6013} &
& {\small 5.2990} &  & {\small 5.5154} &  &  &  & {\small 6.8597} &  &
{\small 5.8902} &  & {\small 6.7664} &  & {\small 7.0454} \\
& {\small (b)} &  & {\small 4.5252} &  &  & {\small 3.8308} &  & {\small %
4.4631} &  & {\small 4.6627} &  &  & {\small 4.1950} &  & {\small 4.8624} &
& {\small 5.0701} &  &  &  & {\small 6.5538} &  & {\small 5.6005} &  &
{\small 6.4589} &  & {\small 6.7333} \\
& {\small (c)} &  & {\small 4.6105} &  &  & {\small 3.9296} &  & {\small %
4.5524} &  & {\small 4.7483} &  &  & {\small 4.3233} &  & {\small 4.9807} &
& {\small 5.1845} &  &  &  & {\small 6.9137} &  & {\small 5.9379} &  &
{\small 6.8196} &  & {\small 7.1005} \\
& {\small (d)} &  & {\small 4.1943} &  &  & {\small 3.5472} &  & {\small %
4.1366} &  & {\small 4.3227} &  &  & {\small 3.8818} &  & {\small 4.5039} &
& {\small 4.6975} &  &  &  & {\small 5.9095} &  & {\small 5.0431} &  &
{\small 5.8237} &  & {\small 6.0730} \\
&  &  &  &  &  &  &  &  &  &  &  &  &  &  &  &  &  &  &  &  &  &  &  &  &  &
&  \\
$^{110}${\small Pd} & {\small (a)} &  & {\small 8.0959} &  &  & {\small %
6.9773} &  & {\small 7.9906} &  & {\small 8.3117} &  &  & {\small 7.7645} &
& {\small 8.8352} &  & {\small 9.1697} &  &  &  & {\small 8.1300} &  &
{\small 6.9989} &  & {\small 8.0220} &  & {\small 8.3467} \\
& {\small (b)} &  & {\small 6.8084} &  &  & {\small 5.8136} &  & {\small %
6.7118} &  & {\small 6.9972} &  &  & {\small 6.4371} &  & {\small 7.3862} &
& {\small 7.6834} &  &  &  & {\small 6.8383} &  & {\small 5.8326} &  &
{\small 6.7394} &  & {\small 7.0279} \\
& {\small (c)} &  & {\small 7.7679} &  &  & {\small 6.6955} &  & {\small %
7.6660} &  & {\small 7.9739} &  &  & {\small 7.4533} &  & {\small 8.4788} &
& {\small 8.7995} &  &  &  & {\small 7.8006} &  & {\small 6.7163} &  &
{\small 7.6961} &  & {\small 8.0074} \\
& {\small (d)} &  & {\small 7.1998} &  &  & {\small 6.1771} &  & {\small %
7.0999} &  & {\small 7.3936} &  &  & {\small 6.8577} &  & {\small 7.8328} &
& {\small 8.1386} &  &  &  & {\small 7.2306} &  & {\small 6.1966} &  &
{\small 7.1283} &  & {\small 7.4252} \\
&  &  &  &  &  &  &  &  &  &  &  &  &  &  &  &  &  &  &  &  &  &  &  &  &  &
&  \\
$^{128}${\small Te} & {\small (a)} &  & {\small 3.4090} &  &  & {\small %
2.9063} &  & {\small 3.3533} &  & {\small 3.4978} &  &  & {\small 3.2453} &
& {\small 3.7190} &  & {\small 3.8698} &  &  &  & {\small 3.4239} &  &
{\small 2.9157} &  & {\small 3.3670} &  & {\small 3.5130} \\
& {\small (b)} &  & {\small 3.7254} &  &  & {\small 3.1201} &  & {\small %
3.6599} &  & {\small 3.8338} &  &  & {\small 3.4477} &  & {\small 4.0197} &
& {\small 4.2012} &  &  &  & {\small 3.7433} &  & {\small 3.1313} &  &
{\small 3.6763} &  & {\small 3.8520} \\
& {\small (c)} &  & {\small 4.0718} &  &  & {\small 3.4962} &  & {\small %
4.0110} &  & {\small 4.1767} &  &  & {\small 3.9073} &  & {\small 4.4529} &
& {\small 4.6257} &  &  &  & {\small 4.0889} &  & {\small 3.5070} &  &
{\small 4.0268} &  & {\small 4.1942} \\
& {\small (d)} &  & {\small 3.9541} &  &  & {\small 3.3394} &  & {\small %
3.8875} &  & {\small 4.0641} &  &  & {\small 3.7048} &  & {\small 4.2856} &
& {\small 4.4700} &  &  &  & {\small 3.9722} &  & {\small 3.3507} &  &
{\small 3.9041} &  & {\small 4.0827} \\
&  &  &  &  &  &  &  &  &  &  &  &  &  &  &  &  &  &  &  &  &  &  &  &  &  &
&  \\
$^{130}${\small Te} & {\small (a)} &  & {\small 4.6291} &  &  & {\small %
4.0296} &  & {\small 4.5729} &  & {\small 4.7458} &  &  & {\small 4.5157} &
& {\small 5.0915} &  & {\small 5.2720} &  &  &  & {\small 4.6472} &  &
{\small 4.0410} &  & {\small 4.5895} &  & {\small 4.7643} \\
& {\small (b)} &  & {\small 3.8628} &  &  & {\small 3.2691} &  & {\small %
3.8033} &  & {\small 3.9741} &  &  & {\small 3.6204} &  & {\small 4.1867} &
& {\small 4.3651} &  &  &  & {\small 3.8805} &  & {\small 3.2802} &  &
{\small 3.8195} &  & {\small 3.9922} \\
& {\small (c)} &  & {\small 4.5565} &  &  & {\small 3.9644} &  & {\small %
4.5007} &  & {\small 4.6716} &  &  & {\small 4.4422} &  & {\small 5.0107} &
& {\small 5.1890} &  &  &  & {\small 4.5744} &  & {\small 3.9756} &  &
{\small 4.5171} &  & {\small 4.6898} \\
& {\small (d)} &  & {\small 3.8553} &  &  & {\small 3.2641} &  & {\small %
3.7959} &  & {\small 3.9660} &  &  & {\small 3.6160} &  & {\small 4.1797} &
& {\small 4.3573} &  &  &  & {\small 3.8728} &  & {\small 3.2751} &  &
{\small 3.8120} &  & {\small 3.9840} \\
&  &  &  &  &  &  &  &  &  &  &  &  &  &  &  &  &  &  &  &  &  &  &  &  &  &
&  \\
$^{150}${\small Nd} & {\small (a)} &  & {\small 3.3454} &  &  & {\small %
2.9312} &  & {\small 3.3122} &  & {\small 3.4317} &  &  & {\small 3.3018} &
& {\small 3.7068} &  & {\small 3.8318} &  &  &  & {\small 3.3583} &  &
{\small 2.9394} &  & {\small 3.3241} &  & {\small 3.4449} \\
& {\small (b)} &  & {\small 2.5216} &  &  & {\small 2.1896} &  & {\small %
2.4937} &  & {\small 2.5894} &  &  & {\small 2.4578} &  & {\small 2.7810} &
& {\small 2.8811} &  &  &  & {\small 2.5319} &  & {\small 2.1961} &  &
{\small 2.5032} &  & {\small 2.5999} \\
& {\small (c)} &  & {\small 3.2683} &  &  & {\small 2.8630} &  & {\small %
3.2358} &  & {\small 3.3528} &  &  & {\small 3.2242} &  & {\small 3.6206} &
& {\small 3.7429} &  &  &  & {\small 3.2809} &  & {\small 2.8710} &  &
{\small 3.2475} &  & {\small 3.3657} \\
& {\small (d)} &  & {\small 2.6013} &  &  & {\small 2.2655} &  & {\small %
2.5735} &  & {\small 2.6703} &  &  & {\small 2.5456} &  & {\small 2.8730} &
& {\small 2.9743} &  &  &  & {\small 2.6117} &  & {\small 2.2722} &  &
{\small 2.5832} &  & {\small 2.6810} \\ \hline\hline
\end{tabular}
\end{table*}

\subsection{Exchange of light Majorana neutrinos}

Employing the HFB wave functions generated with four different parametrizaions of 
effective two-body interaction, the required NTMEs $M^{\left( 0\nu \right) }$ for 
$^{94,96}$Zr, $^{98,100}$Mo, $^{104}$Ru, $^{110}$Pd, $^{128,130}$Te and 
$^{150}$Nd isotopes are calculated with the consideration of two sets of form factors 
and three different parametrizations of the Jastrow type of SRC.   
In Table~\ref{tab1}, the Fermi, Gamow-Teller and tensor components of NTMEs 
$M^{\left(0\nu \right) \text{ }}$ for $^{100}$Mo are presented without 
(HOC - 3rd column) and with the SRC (HOC+SRC -4th to 6th columns) to exhibit 
the role of HOC as well as the SRC explicitly. To test the validity of closure 
approximation used in the present work, the NTMEs $M^{\left( 0\nu \right) }$ are 
also calculated for $\overline{A}/2$ in the energy denominator in the case of 
HOC+SRC given in 7th to 9th columns of Table~\ref{tab1}. It is remarkable that 
the variation in the NTMEs calculated with FNS1 and FNS2 are almost negligible.
 
In Table~\ref{tab2}, we present sets of twenty-four NTMEs $M^{\left( 0\nu \right)
}$ for $^{94,96}$Zr, $^{98,100}$Mo, $^{104}$Ru, $^{110}$Pd, $^{128,130}$Te and
$^{150}$Nd isotopes calculated in the same approximations as mentioned above.
The sets of NTMEs calculated with FNS1 and FNS2 are donated by HOC1 and HOC2,
respectively. It is noticed in general but for $^{128}$Te isotope that the NTMEs
evaluated for both \textit{PQQ}1 and \textit{PQQ}2 parameterizations are quite
close. The inclusion of the hexadecapolar term tends to
reduce them by magnitudes, specifically depending on the structure of  nuclei.
The maximum variation in $M^{\left( 0\nu \right) }$ due to the \textit{PQQHH}1,
\textit{PQQ}2 and \textit{PQQHH}2 parameterizations with respect to \textit{PQQ}1
one lies between 20--25\%. The relative change in NTMEs
$M^{\left( 0\nu \right) }$, by changing the energy denominator to
$\overline{A}/2$ instead of $\overline{A}$ is in between 8.7\%--12.7\%,
which confirms that the dependence
of NTMEs on average excitation energy $\overline{A}$ is small and thus, the
validity of the closure approximation for $\left( \beta ^{-}\beta
^{-}\right) _{0\nu }$ decay is supported.

The study of the radial evolution of NTMEs defined by \cite{simk08} 
\begin{equation}
M^{\left( 0\nu \right) }=\int C^{\left( 0\nu \right) }\left( r\right) dr,
\end{equation}
is the best possible tool to display the role of the HOC as well as the SRC. 
By the study of radial evolution of NTMEs $M^{\left( 0\nu \right) }$, 
\v{S}imkovic $et$ $al.$ in the QRPA \cite{simk08} and Men\'{e}ndez $et$ $al.$ 
in the ISM \cite{mene09} have shown that the magnitude of 
$M^{\left( 0\nu \right) }$ for all
nuclei undergoing $\left( \beta ^{-}\beta ^{-}\right) _{0\nu }$ decay
exhibit a maximum at about the internucleon distance $r\approx 1$ fm, and
that the contributions of decaying pairs coupled to $J=0$ and $J>0$ almost
cancel out beyond $r\approx 3$ fm. Similar observations on the radial
evolution of NTMEs $M^{\left( 0\nu \right) }$ and $M^{\left( 0N\right) }$ 
due to the exchange of light \cite{rath10} and heavy Majorana neutrinos 
\cite{rath12}, respectively have also been reported within the PHFB approach.

\begin{figure}[htbp]
\begin{tabular}{c}
\includegraphics [scale=0.38]{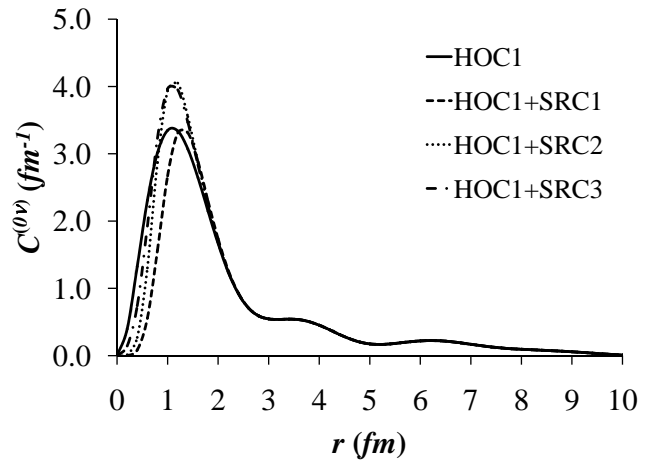} \\
\end{tabular}
\caption{Radial dependence of $C^{(0\nu)}(r)$ for the
$\left( \beta ^{-}\beta ^{-} \right) _{0\nu }$ mode of $^{100}$Mo isotope.}
\label{fig2}
\end{figure}

\begin{figure}[htbp]
\begin{tabular}{c}
\includegraphics [scale=0.32]{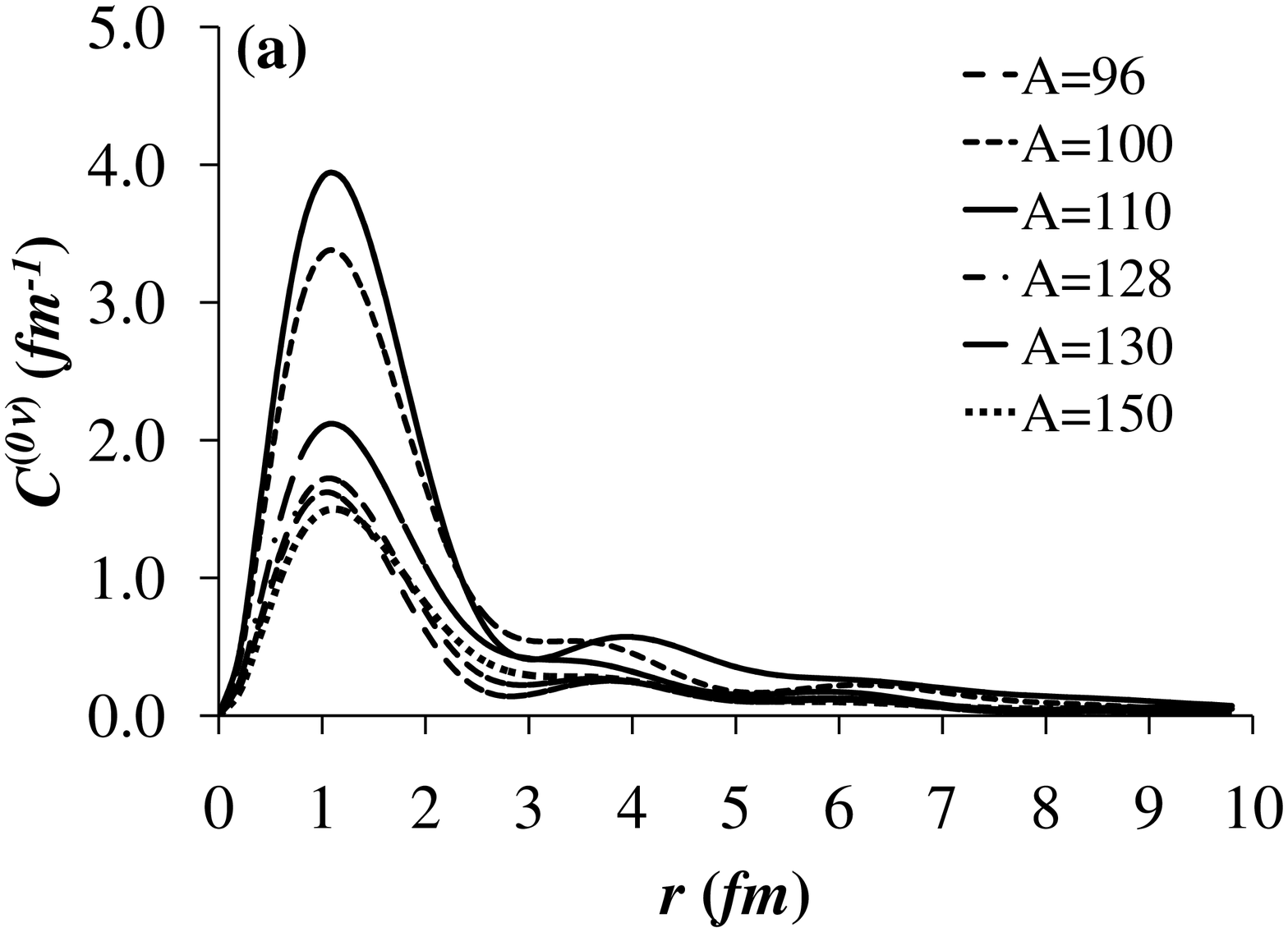} \\
\includegraphics [scale=0.32]{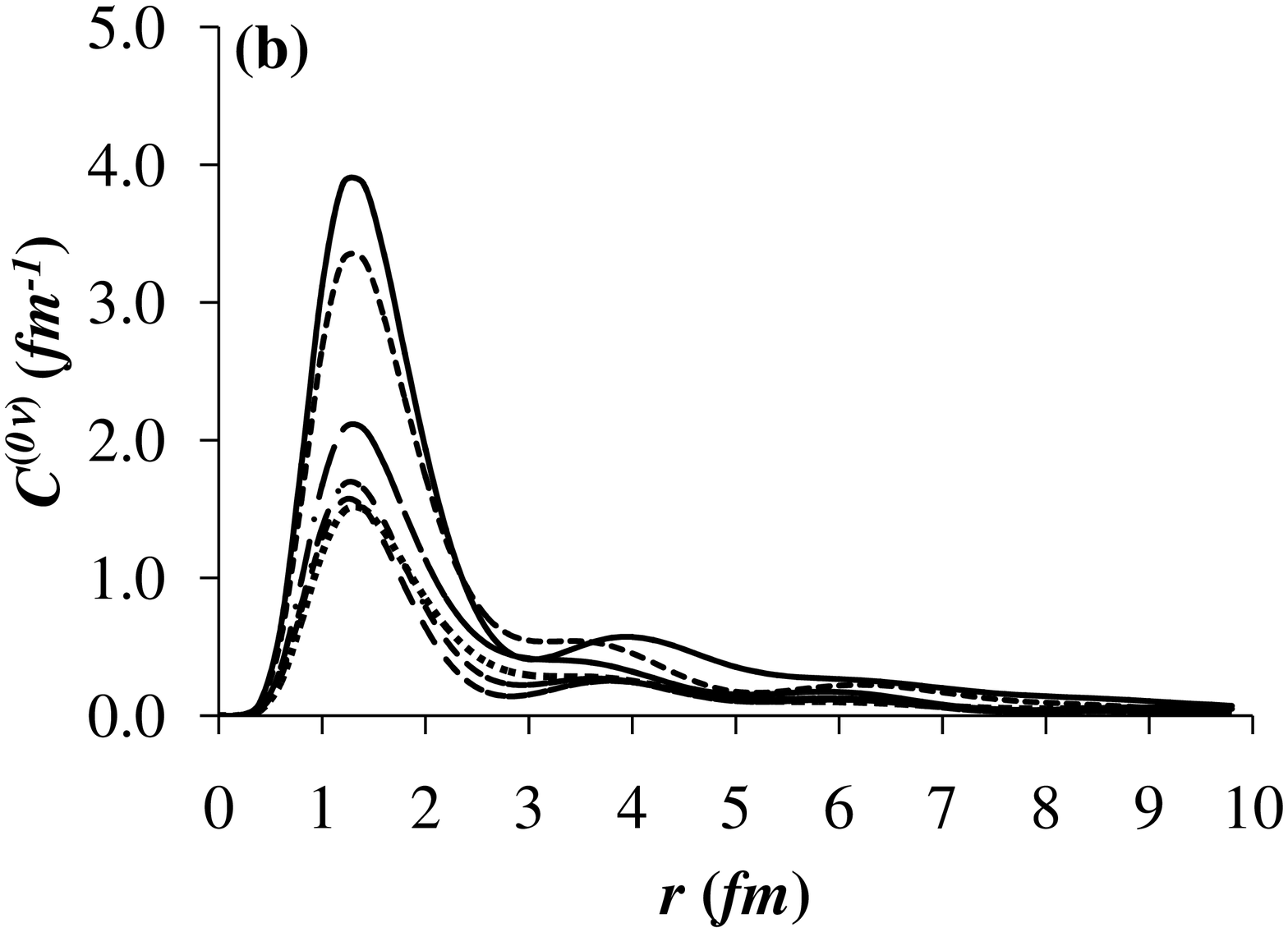} \\
\includegraphics [scale=0.32]{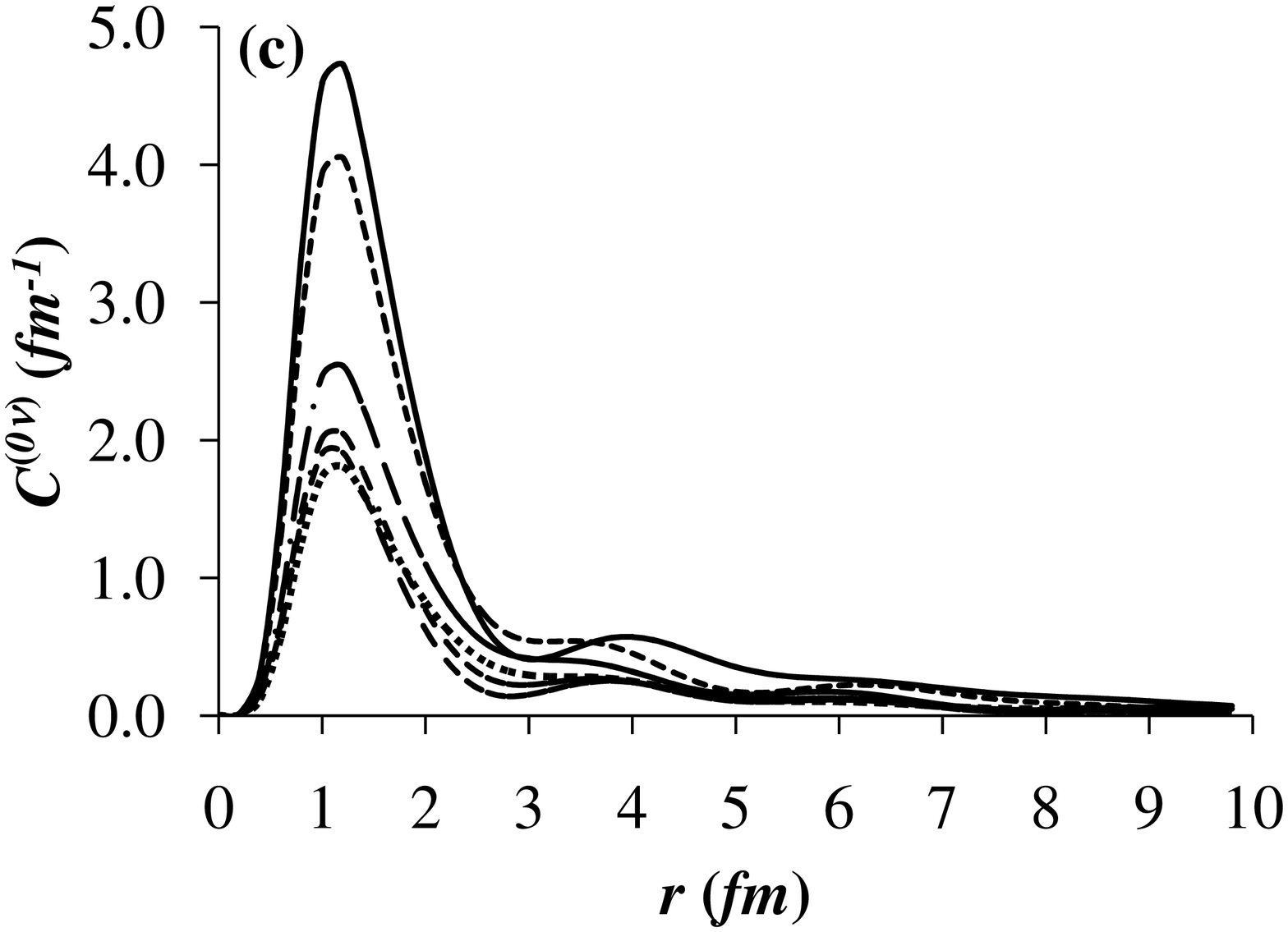} \\
\includegraphics [scale=0.32]{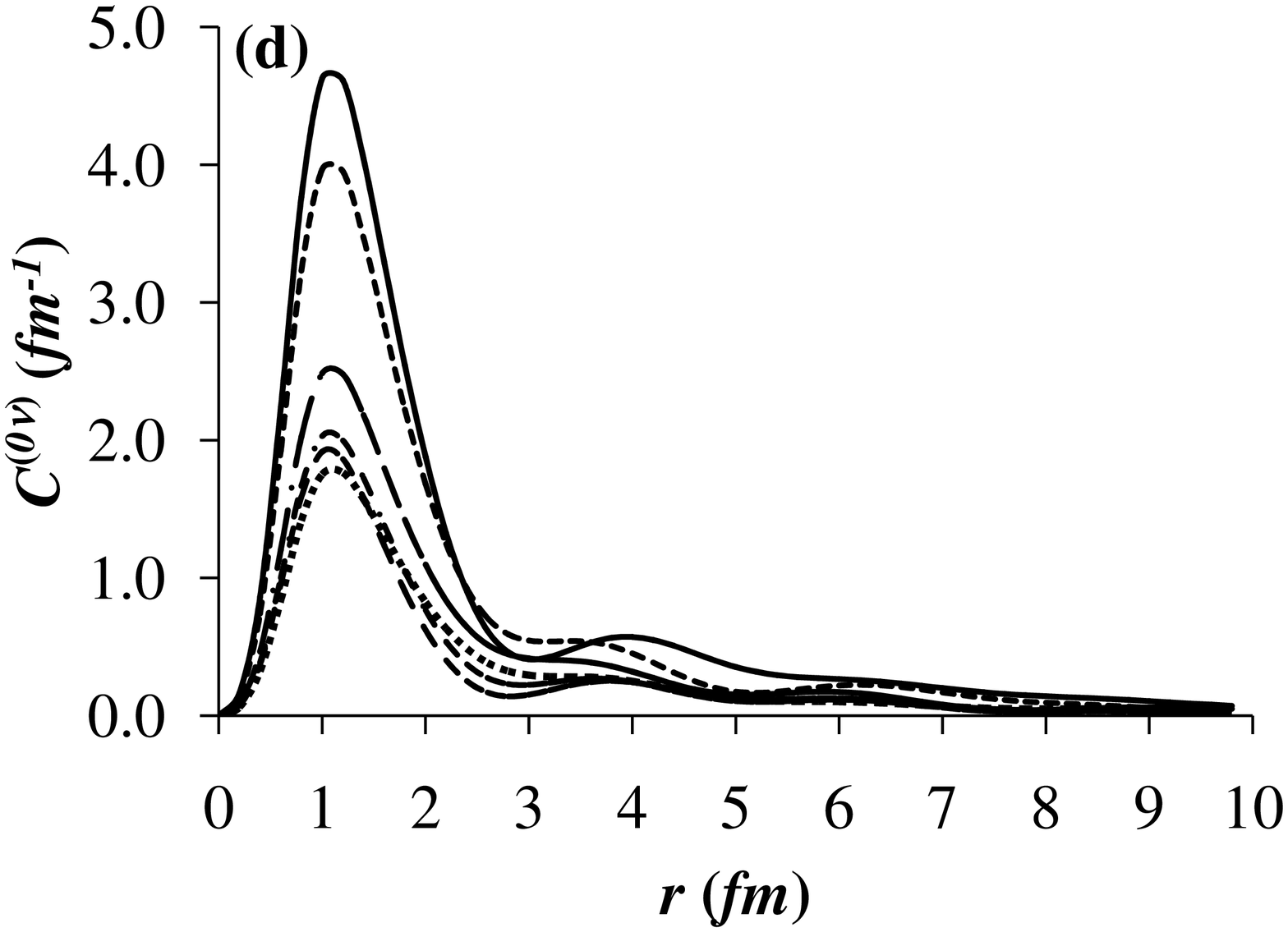} \\
\end{tabular}
\caption{Radial dependence of $C^{(0\nu)}(r)$ for the
$\left( \beta ^{-}\beta ^{-} \right) _{0\nu }$ mode of
$^{96}$Zr, $^{100}$Mo, $^{110}$Pd, $^{128,130}$Te and $^{150}$Nd
isotopes. In this Fig., (a), (b), (c) and (d) correspond to HOC1, HOC1+SRC1,
HOC1+SRC2 and HOC1+SRC3, respectively.}
\label{fig3}
\end{figure}

In Fig. 2, the effects due to the HOC and SRC are made more
transparent by plotting the radial dependence of $C^{\left( 0\nu \right) }$
for $^{100}$Mo isotope with the {\it PQQ}1 parameterization of the effective two
body interaction in four cases, namely HOC1, HOC1+SRC1, HOC1+SRC2 and HOC1+SRC3. 
In Fig. 3, the radial dependence of $C^{\left( 0\nu \right) }$ for the same four
combinations of HOC1 and SRC are plotted for $^{96}$Zr, $^{100}$Mo, $^{110}$%
Pd, $^{128,130}$Te and $^{150}$Nd isotopes. It is noticed that the $%
C^{\left( 0\nu \right) }$ are peaked at $r=1.0$ fm for the HOC1
and the addition of SRC1 and SRC2 shifts the peak to 1.25 fm. However, the
position of the peak remains unchanged at $r=1.0$ fm with the inclusion of
SRC3. Although, the radial distributions of $C^{\left( 0\nu \right) }$
extends up to about 10 fm, the  maximum contribution to the radial evolution 
of  $M^{\left( 0\nu \right) }$ results from the distribution of 
$C^{\left( 0\nu \right) }$ up to 3 fm. In addition, the above observations 
also remain valid with the other
three parameterizations of the effective two-body interaction. The observed
variation in the areas of curves in different cases is
intimately related with the large changes in NTMEs $M^{\left( 0\nu \right) }$
presented in Table~\ref{tab2}. 

\begin{table}[htbp]
\caption{Changes (in \%) of the NTMEs $M^{\left( 0\nu \right) }$
due to exchange of light Majorana neutrinos, for the $\left( \beta ^{-}\beta
^{-}\right) _{0\nu }$ decay with the inclusion of FNS1, HOC1 and HOC1+SRC
(HOC1+SRC1, HOC1+SRC2 and HOC1+SRC3) for the four different parameterizations of
the effective two-body interaction, namely (a) \textit{PQQ}1,
(b) \textit{PQQHH}1, (c) \textit{PQQ}2 and (d) \textit{PQQHH}2.}
\label{tab3}
\begin{tabular}{cccccccccclr}
\hline\hline
&~~~  & FNS1 &  &  & ~~~ & HOC1 &~~~  &  &  &  & HOC1+SRC \\ \hline
(a) &  & 8.94--10.63 &  &  &  & 11.17--12.96 &  &  &~~~  & (i) & 12.38--15.61
\\
&  &  &  &  &  &  &  &  &  & (ii) & 0.99--1.96 \\
&  &  &  &  &  &  &  &  &  & (iii) & 2.48--2.97 \\
(b) &  & 9.40--11.07 &  &  &  & 11.43--13.03 &  &  &  & (i) & 13.17--16.40
\\
&  &  &  &  &  &  &  &  &  & (ii) & 1.11--2.11 \\
&  &  &  &  &  &  &  &  &  & (iii) & 2.50--3.04 \\
(c) &  & 8.95--10.68 &  &  &  & 10.13--13.00 &  &  &  & (i) & 12.40--15.71
\\
&  &  &  &  &  &  &  &  &  & (ii) & 0.99--1.99 \\
&  &  &  &  &  &  &  &  &  & (iii) & 2.39--2.99 \\
(d) &  & 9.25--11.15 &  &  &  & 11.38--12.73 &  &  &  & (i) & 12.91--16.56
\\
&  &  &  &  &  &  &  &  &  & (ii) & 1.07--2.15 \\
&  &  &  &  &  &  &  &  &  & (iii) & 2.51--3.06 \\ \hline\hline
\end{tabular}
\end{table}

In Table~\ref{tab3}, the relative 
changes in NTMEs $M^{\left( 0\nu \right) }$(in \%) due to the different 
approximations are presented. It is noticed that the consideration of FNS1 
induces changes about 9.0\%--11.0\% in
the NTMEs $M_{VV}^{\left( 0\nu \right) }+M_{AA}^{\left( 0\nu \right) }$ of
point nucleon case. The inclusion of HOC1 reduces the
NTMEs further by 11.0\%--13.0\%. The NTMEs $M^{\left( 0\nu \right) }$ are
approximately reduced by 13.0\%--17.0\%, 1.0\%--2.0\% and 2.5\%--3.0\% with the
addition of SRC1, SRC2 and SRC3, respectively, relative to the HOC1 case.

\begin{table}[htbp]
\caption{Deformation ratios $D^{\left( 0\nu \right) }$ of $\left(
\beta ^{-}\beta ^{-}\right) _{0\nu }$ decay for the \textit{PQQ}1 
parameterization.}
\label{tab4}
\begin{tabular}{llccccccc}
\hline\hline
Nuclei~~&~~~~~  & HOC1 &~~~~~~  & \multicolumn{5}{c}{HOC1+SRC} \\ \cline{5-9}
&  & ~~~~ &  & SRC1 &~~~~~~  & SRC2 &~~~~~~  & SRC3 \\ \hline
$^{94}$Zr &  & 2.48 &  & 2.52 &  & 2.49 &  & 2.49 \\
$^{96}$Zr &  & 4.40 &  & 4.53 &  & 4.43 &  & 4.40 \\
$^{98}$Mo &  & 1.94 &  & 1.95 &  & 1.95 &  & 1.94 \\
$^{100}$Mo &  & 2.15 &  & 2.17 &  & 2.15 &  & 2.14 \\
$^{104}$Ru &  & 3.80 &  & 3.91 &  & 3.81 &  & 3.79 \\
$^{110}$Pd &  & 2.61 &  & 2.66 &  & 2.62 &  & 2.61 \\
$^{128}$Te &  & 4.38 &  & 4.50 &  & 4.40 &  & 4.38 \\
$^{130}$Te &  & 2.94 &  & 2.95 &  & 2.94 &  & 2.94 \\
$^{150}$Nd &  & 6.18 &  & 6.17 &  & 6.18 &  & 6.19 \\ \hline\hline
\end{tabular}
\end{table}

\begin{table*}[htbp]
\caption{Average NTMEs $\overline{M}'^{(0\nu)}$ and uncertainties $\Delta
\overline{M}'^{(0\nu)}$ for the $\left( \beta ^{-}\beta ^{-} \right) _{0\nu }$ decay
of $^{94,96}$Zr, $^{98,100}$Mo, $^{104}$Ru, $^{110}$Pd, $^{128,130}$Te and $%
^{150}$Nd isotopes. Both bare and quenched values of $g_{A}$ are considered. Case I
and Case II denote calculations with and without SRC1, respectively. In the 5th column, 
(a), (b), (c), (d) and (e) correspond to Jastrow, FHCh, UCOM and CCM SRC with 
Argonne V18 and CD-Bonn potentials, respectively. The range of values in the 
QRPA and RQRPA for  $g_{A}$=1.0 are tabulated}
\label{tab5}
\begin{tabular}{lllcclcclllllllllcrcccccccc}
\hline\hline
{\small Nuclei} &~~  & {\small g}$_{A}$ &~~  & {\small Case I} &~~  &  & {\small %
Case II} &  &~~  &SRC  &~~  & {\small QRPA} &  &  & {\small RQRPA} &  &  & {\small %
QRPA} &  &~~  & {\small ISM} &  &~~  & {\small EDF} &~~  & {\small IBM} \\
\cline{4-8}
&  &  &  & $\overline{M}^{(0\nu )}$ &  &  & $\overline{M}^{(0\nu )}$ &  &  &
&  & {\small \cite{faes12}} &  &  & {\small \cite{faes12}} &  &  & {\small
\cite{kor07a,suho10,suho11}} &  &  & {\small \cite{caur08,mene09}} &  &  &
{\small \cite{rodr10}} &  & {\small \cite{bare13}} \\ \hline
$^{94}${\small Zr} &  & {\small 1.254} &  & \multicolumn{1}{l}{{\small 3.88}$%
\pm ${\small 0.37}} &  &  & \multicolumn{1}{l}{{\small 4.08}$\pm ${\small %
0.24}} &  &  & {\small (c)} &  &  &  &  &  &  &  & {\small 3.52}$\pm 0.39$ &
&  &  &  &  &  &  &  \\
&  & {\small 1.0} &  & \multicolumn{1}{l}{{\small 2.76}$\pm 0.26$} &  &  &
\multicolumn{1}{l}{{\small 2.90}$\pm ${\small 0.17}} &  &  & {\small (c)} &
&  &  &  &  &  &  & {\small 3.90}$\pm 0.13$ &  &  &  &  &  &  &  &  \\
$^{96}${\small Zr} &  & {\small 1.254} &  & \multicolumn{1}{l}{{\small 2.86}$%
\pm ${\small 0.26}} &  &  & \multicolumn{1}{l}{{\small 3.03}$\pm ${\small %
0.12}} &  &  & {\small (a)} &  & {\small 1.22}$\pm 0.03$ &  &  & {\small 1.31%
}$\pm 0.15$ &  &  &  &  &  &  &  &  &  &  & {\small 2.53} \\
&  &  &  &  &  &  &  &  &  & {\small (b)} &  & {\small 1.23}$\pm 0.04$ &  &
& {\small 1.33}$\pm 0.15$ &  &  &  &  &  &  &  &  &  &  &  \\
&  &  &  &  &  &  &  &  &  & {\small (c)} &  & {\small 1.77}$\pm 0.02$ &  &
& {\small 1.77}$\pm 0.02$ &  &  & {\small 3.117} &  &  &  &  &  & {\small %
5.65} &  &  \\
&  &  &  &  &  &  &  &  &  & {\small (d)} &  & {\small 2.07}$\pm 0.10$ &  &
& {\small 2.01}$\pm 0.17$ &  &  &  &  &  &  &  &  &  &  & {\small 2.89} \\
&  &  &  &  &  &  &  &  &  & {\small (e)} &  & {\small 2.28}$\pm 0.03$ &  &
& {\small 2.19}$\pm 0.22$ &  &  &  &  &  &  &  &  &  &  & {\small 3.00} \\
&  & {\small 1.0} &  & \multicolumn{1}{l}{{\small 2.04}$\pm 0.19$} &  &  &
\multicolumn{1}{l}{{\small 2.16}$\pm ${\small 0.09}} &  &  &  &  & {\small %
1.32}$\pm 0.08-$ &  &  & {\small 1.22}$\pm 0.12-$ &  &  &  &  &  &  &  &  &
&  &  \\
&  &  &  &  &  &  &  &  &  &  &  & $2.11\pm 0.12$ &  &  & $1.88\pm 0.16$ &
&  & {\small 2.764} &  &  &  &  &  &  &  &  \\
$^{98}${\small Mo} &  & {\small 1.254} &  & \multicolumn{1}{l}{{\small 6.49}$%
\pm ${\small 0.55}} &  &  & \multicolumn{1}{l}{{\small 6.81}$\pm ${\small %
0.32}} &  &  &  &  &  &  &  &  &  &  &  &  &  &  &  &  &  &  &  \\
&  & {\small 1.0} &  & \multicolumn{1}{l}{{\small 4.64}$\pm 0.41$} &  &  &
\multicolumn{1}{l}{{\small 4.88}$\pm ${\small 0.25}} &  &  &  &  &  &  &  &
&  &  &  &  &  &  &  &  &  &  &  \\
$^{100}${\small Mo} &  & {\small 1.254} &  & \multicolumn{1}{l}{{\small 6.26}%
$\pm ${\small 0.63}} &  &  & \multicolumn{1}{l}{{\small 6.59}$\pm ${\small %
0.44}} &  &  & {\small (a)} &  & {\small 3.64}$\pm 0.21$ &  &  & {\small 3.03%
}$\pm 0.21$ &  &  &  &  &  &  &  &  &  &  & {\small 3.73} \\
&  &  &  &  &  &  &  &  &  & {\small (b)} &  & {\small 3.73}$\pm 0.21$ &  &
& {\small 3.12}$\pm 0.21$ &  &  &  &  &  &  &  &  &  &  &  \\
&  &  &  &  &  &  &  &  &  & {\small (c)} &  & {\small 4.71}$\pm 0.28$ &  &
& {\small 3.88}$\pm 0.26$ &  &  & {\small 3.931} &  &  &  &  &  & {\small %
5.08} &  &  \\
&  &  &  &  &  &  &  &  &  & {\small (d)} &  & {\small 5.18}$\pm 0.36$ &  &
& {\small 4.20}$\pm 0.24$ &  &  &  &  &  &  &  &  &  &  & {\small 4.31} \\
&  &  &  &  &  &  &  &  &  & {\small (e)} &  & {\small 5.73}$\pm 0.34$ &  &
& {\small 4.67}$\pm 0.31$ &  &  &  &  &  &  &  &  &  &  & {\small 4.50} \\
&  & {\small 1.0} &  & \multicolumn{1}{l}{{\small 4.49}$\pm 0.47$} &  &  &
\multicolumn{1}{l}{{\small 4.73}$\pm ${\small 0.33}} &  &  &  &  & {\small %
2.96}$\pm 0.15-$ &  &  & {\small 2.55}$\pm 0.13-$ &  &  & {\small 3.103} &
&  &  &  &  &  &  &  \\
&  &  &  &  &  &  &  &  &  &  &  & $4.44\pm 0.24$ &  &  & $3.75\pm 0.21$ &
&  &  &  &  &  &  &  &  &  &  \\
$^{104}${\small Ru} &  & {\small 1.254} &  & \multicolumn{1}{l}{{\small 4.36}%
$\pm ${\small 0.44}} &  &  & \multicolumn{1}{l}{{\small 4.61}$\pm ${\small %
0.28}} &  &  & {\small (c)} &  &  &  &  &  &  &  & {\small 3.73}$\pm 1.56$ &
&  &  &  &  &  &  &  \\
&  & {\small 1.0} &  & \multicolumn{1}{l}{{\small 3.15}$\pm 0.33$} &  &  &
\multicolumn{1}{l}{{\small 3.33}$\pm ${\small 0.22}} &  &  & {\small (c)} &
&  &  &  &  &  &  & {\small 4.10}$\pm 1.40$ &  &  &  &  &  &  &  &  \\
$^{110}${\small Pd} &  & {\small 1.254} &  & \multicolumn{1}{l}{{\small 7.16}%
$\pm ${\small 0.74}} &  &  & \multicolumn{1}{l}{{\small 7.53}$\pm ${\small %
0.54}} &  &  & {\small (a)} &  &  &  &  &  &  &  &  &  &  &  &  &  &  &  &
{\small 3.62} \\
&  &  &  &  &  &  &  &  &  & {\small (c)} &  &  &  &  &  &  &  & {\small 5.63%
}$\pm 0.49$ &  &  &  &  &  &  &  &  \\
&  &  &  &  &  &  &  &  &  & {\small (d)} &  &  &  &  &  &  &  &  &  &  &  &
&  &  &  & {\small 4.15} \\
&  &  &  &  &  &  &  &  &  & {\small (e)} &  &  &  &  &  &  &  &  &  &  &  &
&  &  &  & {\small 4.31} \\
&  & {\small 1.0} &  & \multicolumn{1}{l}{{\small 5.12}$\pm 0.55$} &  &  &
\multicolumn{1}{l}{{\small 5.39}$\pm ${\small 0.41}} &  &  & {\small (c)} &
&  &  &  &  &  &  & {\small 5.14}$\pm 0.69$ &  &  &  &  &  &  &  &  \\
$^{128}${\small Te} &  & {\small 1.254} &  & \multicolumn{1}{l}{{\small 3.62}%
$\pm ${\small 0.39}} &  &  & \multicolumn{1}{l}{{\small 3.82}$\pm ${\small %
0.28}} &  &  & {\small (a)} &  & {\small 3.97}$\pm 0.14$ &  &  & {\small 3.52%
}$\pm 0.13$ &  &  &  &  &  & {\small 2.26} &  &  &  &  & {\small 4.48} \\
&  &  &  &  &  &  &  &  &  & {\small (b)} &  & {\small 4.15}$\pm 0.15$ &  &
& {\small 3.68}$\pm 0.14$ &  &  &  &  &  &  &  &  &  &  &  \\
&  &  &  &  &  &  &  &  &  & {\small (c)} &  & {\small 5.04}$\pm 0.15$ &  &
& {\small 4.45}$\pm 0.15$ &  &  & {\small 5.62} &  &  & {\small 2.88} &  &
& {\small 4.11} &  &  \\
&  &  &  &  &  &  &  &  &  & {\small (d)} &  & {\small 5.38}$\pm 0.17$ &  &
& {\small 4.71}$\pm 0.17$ &  &  &  &  &  &  &  &  &  &  & {\small 4.97} \\
&  &  &  &  &  &  &  &  &  & {\small (e)} &  & {\small 5.99}$\pm 0.17$ &  &
& {\small 5.26}$\pm 0.16$ &  &  &  &  &  &  &  &  &  &  & {\small 5.13} \\
&  & {\small 1.0} &  & \multicolumn{1}{l}{{\small 2.61}$\pm 0.28$} &  &  &
\multicolumn{1}{l}{{\small 2.75}$\pm ${\small 0.20}} &  &  &  &  & {\small %
3.11}$\pm 0.09-$ &  &  & {\small 2.77}$\pm 0.09-$ &  &  & {\small 3.74} &  &
&  &  &  &  &  &  \\
&  &  &  &  &  &  &  &  &  &  &  & $4.54\pm 0.13$ &  &  & $4.0\pm 0.12$ &  &
&  &  &  &  &  &  &  &  &  \\
$^{130}${\small Te} &  & {\small 1.254} &  & \multicolumn{1}{l}{{\small 4.05}%
$\pm ${\small 0.49}} &  &  & \multicolumn{1}{l}{{\small 4.26}$\pm ${\small %
0.39}} &  &  & {\small (a)} &  & {\small 3.56}$\pm 0.13$ &  &  & {\small 3.22%
}$\pm 0.13$ &  &  &  &  &  & {\small 2.04} &  &  &  &  & {\small 4.03} \\
&  &  &  &  &  &  &  &  &  & {\small (b)} &  & {\small 3.72}$\pm 0.14$ &  &
& {\small 3.36}$\pm 0.15$ &  &  &  &  &  &  &  &  &  &  &  \\
&  &  &  &  &  &  &  &  &  & {\small (c)} &  & {\small 4.53}$\pm 0.12$ &  &
& {\small 4.07}$\pm 0.13$ &  &  & {\small 5.12} &  &  & {\small 2.65} &  &
& {\small 5.13} &  &  \\
&  &  &  &  &  &  &  &  &  & {\small (d)} &  & {\small 4.77}$\pm 0.15$ &  &
& {\small 4.27}$\pm 0.15$ &  &  &  &  &  &  &  &  &  &  & {\small 4.47} \\
&  &  &  &  &  &  &  &  &  & {\small (e)} &  & {\small 5.37}$\pm 0.13$ &  &
& {\small 4.80}$\pm 0.14$ &  &  &  &  &  &  &  &  &  &  & {\small 4.61} \\
&  & {\small 1.0} &  & \multicolumn{1}{l}{{\small 2.91}$\pm 0.35$} &  &  &
\multicolumn{1}{l}{{\small 3.07}$\pm ${\small 0.28}} &  &  &  &  & {\small %
2.55}$\pm 0.08-$ &  &  & {\small 2.15}$\pm 0.14-$ &  &  & {\small 3.48} &  &
&  &  &  &  &  &  \\
&  &  &  &  &  &  &  &  &  &  &  & $4.11\pm 0.08$ &  &  & $3.69\pm 0.09$ &
&  &  &  &  &  &  &  &  &  &  \\
$^{150}${\small Nd} &  & {\small 1.254} &  & \multicolumn{1}{l}{{\small 2.83}%
$\pm ${\small 0.42}} &  &  & \multicolumn{1}{l}{{\small 2.96}$\pm ${\small %
0.39}} &  &  & {\small (a)} &  & {\small 2.95}$\pm 0.04$ &  &  &  &  &  &  &
&  &  &  &  &  &  & {\small 2.32} \\
&  &  &  &  &  &  &  &  &  & {\small (c)} &  &  &  &  &  &  &  &  &  &  &  &
&  & {\small 1.71} &  &  \\
&  &  &  &  &  &  &  &  &  & {\small (d)} &  &  &  &  &  &  &  &  &  &  &  &
&  &  &  & {\small 2.74} \\
&  &  &  &  &  &  &  &  &  & {\small (e)} &  &  &  &  &  &  &  &  &  &  &  &
&  &  &  & {\small 2.88} \\
&  & {\small 1.0} &  & \multicolumn{1}{l}{{\small 2.04}$\pm 0.31$} &  &  &
\multicolumn{1}{l}{{\small 2.13}$\pm ${\small 0.29}} &  &  &  &  &  &  &  &
&  &  &  &  &  &  &  &  &  &  &  \\ \hline\hline
\end{tabular}
\end{table*}

The effect of deformation on $M^{\left( 0\nu \right) }$ is quantified by the
quantity $D^{\left( 0\nu \right) }$ defined as the ratio of $M^{\left( 0\nu
\right) }$ at zero deformation ($\zeta _{qq}=0$) and full deformation ($%
\zeta _{qq}=1$) \cite{chat08}. 
\begin{equation}
D^{\left( 0\nu \right) }=\frac{M^{\left( 0\nu \right) }(\zeta _{qq}=0)}{%
M^{\left( 0\nu \right) }(\zeta _{qq}=1)},
\end{equation}
In Table~\ref{tab4}, we tabulate the values of $D^{\left( 0\nu \right) }$
for $^{94,96}$Zr, $^{98,100}$Mo, $^{104}$Ru, $^{110}$Pd, $^{128,130}$Te and $%
^{150}$Nd nuclei. It is observed that owing to deformation effects, the NTMEs 
$M^{\left( 0\nu \right) }$ are suppressed by factor of about 2--6 in the mass
range $A=90-150$. Thus, the deformation plays a crucial role in the nuclear 
structure aspects of $\left(\beta ^{-}\beta ^{-}\right) _{0\nu }$ decay.

\begin{table*}[htbp]
\caption{Average figure of merits $\overline{C}_{mm}^{(0\nu)}$ and uncertainties $\Delta
\overline{C}_{mm}^{(0\nu)}$ for the $\left( \beta ^{-}\beta ^{-} \right) _{0\nu }$ decay
of $^{94,96}$Zr, $^{98,100}$Mo, $^{104}$Ru, $^{110}$Pd, $^{128,130}$Te and $%
^{150}$Nd isotopes. Both bare and quenched values of $g_{A}$ are considered. Case II denote 
calculations without SRC1.}
\label{tab6}
\begin{tabular}{llllllllcclr}
\hline\hline
{\small Nuclei} & ~~~{\small g}$_{A}$ &~~~~~~~~  & {\small Case II} &~~~~~~  & ~~~{\small Exp. }$%
T_{1/2}^{\left( 0\nu \right) }${\small \ (yr}$)$ & {\small Ref.}~~~~ & $%
\left\langle m_{\nu }\right\rangle ${\small \ (eV)} &~~~~~~  & $T_{1/2}^{\left(
0\nu \right) }${\small \ (}$m_{\nu }=0.05${\small \ eV)} &~~~~~~  & $\xi ^{\left(
0\nu \right) }$ \\ \cline{4-5}
&  &  & $\overline{C}_{mm}^{(0\nu )}$ &~~~$\overline{\Delta C}_{mm}^{(0\nu )}$
&  &  &  &  &  &  &  \\ \hline
$^{94}${\small Zr} & ~~~{\small 1.254} &  & {\small 0.1076} &~~~{\small 0.0125} &
~~~{\small 1.9}$\times 10^{19}$ & {\small \cite{arno99}}~~~~ & {\small 6.99}$\times
10^{2}$ &  & {\small 3.72}$_{-0.39}^{+0.49}\times 10^{27}$ &  & {\small 16.74%
} \\
& ~~~{\small 1.0} &  & {\small 0.0545} &~~~{\small 0.0062} &  &  & {\small 9.83}$%
\times 10^{2}$ &  & {\small 7.34}$_{-0.76}^{+0.95}\times 10^{27}$ &  &
{\small 11.91} \\
$^{96}${\small Zr} & ~~~{\small 1.254} &  & {\small 2.0851} &~~~{\small 0.1592} &
~~~{\small 9.2}$\times 10^{21}$ & {\small \cite{argy10}}~~~~ & {\small 7.22} &  &
{\small 1.92}$_{-0.14}^{+0.16}\times 10^{26}$ &  & {\small 73.74} \\
& ~~~{\small 1.0} &  & {\small 1.0658} &~~~{\small 0.0856} &  &  & {\small 10.10}
&  & {\small 3.75}$_{-0.28}^{+0.33}\times 10^{26}$ &  & {\small 52.72} \\
$^{98}${\small Mo} & ~~~{\small 1.254} &  & {\small 0.0032} &~~~{\small 0.0003} &
~~~{\small 1.0}$\times 10^{14}$ & {\small \cite{frem52}}~~~~ & {\small 1.78}$\times
10^{6}$ &  & {\small 1.26}$_{-0.11}^{+0.13}\times 10^{29}$ &  & {\small 2.87}
\\
& ~~~{\small 1.0} &  & {\small 0.0016} &~~~{\small 0.0002} &  &  & {\small 2.48}$%
\times 10^{6}$ &  & {\small 2.46}$_{-0.23}^{+0.28}\times 10^{29}$ &  &
{\small 2.06} \\
$^{100}${\small Mo} & ~~~{\small 1.254} &  & {\small 7.7464} &~~~{\small 1.0044}
& ~~~{\small 1.0}$\times 10^{24}$ & {\small \cite{tret11}}~~~~ & {\small 0.36} &  &
{\small 5.16}$_{-0.59}^{+0.77}\times 10^{25}$ &  & {\small 141.9} \\
& ~~~{\small 1.0} &  & {\small 3.9874} &~~~{\small 0.5466} &  &  & {\small 0.50}
&  & {\small 1.00}$_{-0.12}^{+0.16}\times 10^{26}$ &  & {\small 101.8} \\
$^{104}${\small Ru} & ~~~{\small 1.254} &  & {\small 0.2594} &~~~{\small 0.0318}
& ~~~ & ~~~~ &  
 &  & {\small 1.54}$_{-0.17}^{+0.22}\times 10^{27}$ &  &
{\small 25.98} \\
& ~~~{\small 1.0} &  & {\small 0.1354} &~~~{\small 0.0180} &  &  & 
 &  & {\small 2.96}$_{-0.35}^{+0.45}\times 10^{27}$ &  &
{\small 18.76} \\
$^{110}${\small Pd} & ~~~{\small 1.254} &  & {\small 3.1058} &~~~{\small 0.4446}
& ~~~{\small 6.0}$\times 10^{17}$ & {\small \cite{wint52}}~~~~ & {\small 7.33}$%
\times 10^{2}$ &  & {\small 1.29}$_{-0.16}^{+0.22}\times 10^{26}$ &  &
{\small 89.84} \\
& ~~~{\small 1.0} &  & {\small 1.5904} &~~~{\small 0.2398} &  &  & {\small 1.02}$%
\times 10^{3}$ &  & {\small 2.52}$_{-0.33}^{+0.45}\times 10^{26}$ &  &
{\small 64.27} \\
$^{128}${\small Te} & ~~~{\small 1.254} &  & {\small 0.1038} &~~~{\small 0.0148}
& ~~~{\small 1.1}$\times 10^{23}$ & {\small \cite{arna03}}~~~~ & {\small 9.36} &  &
{\small 3.86}$_{-0.48}^{+0.64}\times 10^{27}$ &  & {\small 16.42} \\
& ~~~{\small 1.0} &  & {\small 0.0539} &~~~{\small 0.0076} &  &  & {\small 12.98}
&  & {\small 7.42}$_{-0.92}^{+1.22}\times 10^{27}$ &  & {\small 11.84} \\
$^{130}${\small Te} & ~~~{\small 1.254} &  & {\small 3.1488} &~~~{\small 0.5757}
& ~~~{\small 3.0}$\times 10^{24}$ & {\small \cite{arna08}}~~~~ & {\small 0.32} &  &
{\small 1.27}$_{-0.20}^{+0.28}\times 10^{26}$ &  & {\small 90.32} \\
& ~~~{\small 1.0} &  & {\small 1.6295} &~~~{\small 0.2994} &  &  & {\small 0.45}
&  & {\small 2.46}$_{-0.38}^{+0.55}\times 10^{26}$ &  & {\small 64.97} \\
$^{150}${\small Nd} & ~~~{\small 1.254} &  & {\small 7.2330} &~~~{\small 1.8979}
& ~~~{\small 1.8}$\times 10^{22}$ & {\small \cite{argy09}}~~~~ & {\small 2.77} &  &
{\small 5.53}$_{-1.15}^{+1.97}\times 10^{25}$ &  & {\small 136.3} \\
& ~~~{\small 1.0} &  & {\small 3.7480} &~~~{\small 0.9919} &  &  & {\small 3.85}
&  & {\small 1.07}$_{-0.22}^{+0.38}\times 10^{26}$ &  & {\small 98.10} \\
\hline
\end{tabular}
\end{table*}

In the present statistical analysis, we employ only the twenty-four NTMEs listed 
in the columns 4--6 (HOC1+SRC) and columns 11--13 of Table~\ref{tab2}, employing 
the bare and quenched values of axial vector coupling constant $g_{A}=1.254$ and $%
g_{A}=1.0$, respectively. In Table~\ref{tab5}, we display the calculated 
averages $\overline{M}^{(0\nu )}$ and their variances $\Delta \overline{M}^{(0\nu )}$ 
for $^{94,96}$Zr, $^{98,100}$Mo, $^{110}$Pd, $^{128,130}$Te
and $^{150}$Nd isotopes isotopes along with all the available theoretical results 
in other models.
It turns out that the uncertainties $\Delta \overline{M}^{(0\nu )}$
are of the order of 10\%, but for $^{130}$Te and $^{150}$Nd isotopes for
which $\Delta \overline{M}^{(0\nu )}$ are approximately 12 \% and 15\%,
respectively. Further, the effect due to the Miller-Spenser parameterization
of Jastrow type of SRC is estimated by evaluating the same mean $\overline{M}%
^{(0\nu )}$ and their standard deviations $\Delta \overline{M}^{(0\nu )}$
for sixteen NTMEs calculated using SRC2 and SRC3 parameterizations. By excluding 
the NTMEs due to SRC1 in the statistical analysis, the uncertainties reduce by 
1.5\%--5\%.  

The figure-of-merit defined by
\begin{equation}
C_{mm}^{\left( 0\nu \right) }=\frac{10^{24}\times G_{01}}{m_{e}^{2}}\left| M^{(0\nu
)}\right| ^{2},
\end{equation}
is another NTME related convenient quantity and is usually used in the
analysis of experimental data. The arbitrary scaling factor $10^{24}$ is
used so that the $C_{mm}^{\left( 0\nu \right) }$ are of the order of unity.
Recently, the phase space factors have been calculated by Kotila and Iachello 
\cite{koti12} with screening correction. However, we evaluate the twenty-four 
$C_{mm}^{\left( 0\nu \right) }$ using rescaled phase
space factors of Boehm and Vogel \cite{boeh92} for $g_{A}=1.254$ and perform
a statistical analysis for estimating the averages $\overline{C}%
_{mm}^{\left( 0\nu \right) }$ and uncertainties $\Delta \overline{C}%
_{mm}^{\left( 0\nu \right) }$. In the 3rd and 4rth columns of 
Table ~\ref{tab6}, the averages $\overline{C}_{mm}^{\left( 0\nu \right) }$ and 
uncertainties $\Delta
\overline{C}_{mm}^{\left( 0\nu \right) }$ are displayed. The uncertainties in
$\Delta \overline{C}_{mm}^{\left( 0\nu \right) }$ are about twice of those
of  $\Delta \overline{M}^{(0\nu )}$.  

The limits on the effective mass of light neutrino $\left\langle
m_{\nu }\right\rangle $ are extracted from the largest observed limits on
half-lives $T_{1/2}^{0\nu }$ of $\left( \beta ^{-}\beta ^{-}\right) _{0\nu }$
decay using the average $\overline{C}_{mm}^{\left( 0\nu \right) }$. It is
observed that the extracted limits on $\left\langle m_{\nu }\right\rangle $
for $^{100}$Mo and $^{130}$Te nuclei are $0.36$ eV--$0.50$ eV and $0.32$ eV--$0.45$ eV, 
respectively. In the last but one
column of Table~\ref{tab6}, the predicted half-lives of $\left( \beta ^{-}\beta
^{-}\right) _{0\nu }$ decay of $^{94,96}$Zr, $^{98,100}$Mo, $^{110}$Pd, $%
^{128,130}$Te and $^{150}$Nd isotopes are given for $\left\langle m_{\nu
}\right\rangle =50$ meV. In the last Column of Table~\ref{tab6}, we present the
nuclear sensitivities, which are related to mass sensitivities, defined by
\cite{simk99}
\begin{equation}
\xi ^{\left( 0\nu \right) }=10^{8}\sqrt{G_{01}}\left| M^{(0\nu )}\right|,
\end{equation}
with an arbitrary normalization factor 10$^{8}$ so that the nuclear
sensitivities turn out to be order of unity. It is observed that the nuclear
sensitivities for $\left( \beta ^{-}\beta ^{-}\right) _{0\nu }$ decay of $%
^{100}$Mo, $^{150}$Nd, $^{130}$Te, $^{110}$Pd, $^{96}$Zr, $^{104}$Ru, 
$^{94}$Zr $^{128}$Te,and $^{98}$Mo isotopes are in the decreasing order
of their magnitudes.

\subsection{Majoron emission}

In the classical Majoron model, the NTMEs $M^{\left( 0\nu \phi \right) }$
for the $\left( \beta ^{-}\beta ^{-}\phi\right) _{0\nu }$ decay are same as 
the $M^{\left( 0\nu \right) }$ for $\left( \beta ^{-}\beta ^{-}\right) _{0\nu }$ decay. 
Hence, the average NTMEs $%
\overline{M}^{(0\nu \phi )}$ and uncertainties $\Delta \overline{M}^{(0\nu
\phi )}$ are same as the $\overline{M}^{(0\nu )}$ and $\Delta \overline{M}%
^{(0\nu )}$, respectively. 
The phase space factors $G_{0M}$ for the 0$^{+}\rightarrow $0$^{+}$ transition 
of $\left(\beta ^{-} \beta ^{-} \phi \right) _{0\nu }$ decay mode have been calculated 
by Hirsch \textit{et al.} \cite{himj96} using $\ g_{A}$= 1.25. 
We calculate the phase space factors given in Table~\ref{tab7} for the $0^{+}\to0^{+}$ 
transition of $^{94,96}$Zr, $^{98,100}$Mo, $^{104}$Ru and $\,^{110}$Pd, $^{128,130}$Te 
and $^{150}$Nd isotopes with $g_{A}$= 1.254. The maximum difference between 
the two sets of phase space factors is less than 5 \%.

The extracted limits on the effective Majoron-neutrino
coupling parameter $\left\langle g_{M}\right\rangle $ form the largest
limits on the observed half-lives $T_{1/2}^{\left( 0\nu \phi \right) }$ are given in
Table~\ref{tab7}. The most stringent extracted limit on $\left\langle
g_{M}\right\rangle < \left( 2.22-3.09\right) \times 10^{-5}$. In the column 6
of the same Table~\ref{tab7}, the predicted half-lives $T_{1/2}^{\left( 0\nu \phi
\right) }$ for $\left\langle g_{M}\right\rangle =10^{-6}$ are displayed. The
sensitivities $\xi ^{\left( 0\nu \phi \right) }$ defined similar to $\xi
^{\left( 0\nu \right) }$ are presented in the last column of the Table~\ref{tab7}. 
It is noticed that the sensitivities for $\left( \beta ^{-}\beta ^{-}\phi
\right) _{0\nu }$ decay mode of $^{98}$Mo, $^{128}$Te, $^{94}$Zr, $^{104}$Ru, 
$^{110}$Pd, $^{130}$Te, $^{96}$Zr, $^{100}$Mo and $^{150}$Nd isotopes
are in the increasing order of their magnitudes.

\begin{table*}[htbp]
\caption{Phase space factors, extracted effective Majoron-neutrino coupling 
 $\left\langle g_{M}\right\rangle $, predicted half-lives T$_{1/2}^{(0\nu \phi )}$
and sensitivities for the $\left( \beta ^{-}\beta ^{-} \phi \right) _{0\nu }$ decay
of $^{94,96}$Zr, $^{98,100}$Mo, $^{104}$Ru, $^{110}$Pd, $^{128,130}$Te and $%
^{150}$Nd isotopes.}
\label{tab7}
\begin{tabular}{ccccccccccccr}
\hline\hline
{\small Nuclei} &~~~~~  & $G_{0M}$ &~~~~~~  & {\small Expt. T}$_{1/2}^{(0\nu \phi )}$%
{\small \ (yr)} & {\small Ref. }~~~~ & $g_A$  &~~~~~~  & $\left\langle g_{M}\right\rangle $
& ~~~~~~ & $
\begin{array}{c}
Predicted\text{ }T_{1/2}^{(0\nu \phi )}\text{ }(yr) \\
\left\langle g_{M}\right\rangle =10^{-6}
\end{array}
$ &  & $\xi ^{\left( 0\nu \phi \right) }$ \\ \hline
&  &  &  &  &  &  &  &  &  &  &  &  \\
$^{94}${\small Zr} &  & {\small 1.299}$\times 10^{-17}$ &  & {\small 2.3}$%
\times 10^{18}$ & \cite{arno99}~~~~ & {\small (a)} &  & {\small 4.49}$\times
10^{-2}$ &  & {\small 4.63}$_{-0.50}^{+0.59}\times 10^{27}$ &  & {\small 1.47%
} \\
&  &  &  &  &  & {\small (b)} &  & {\small 6.31}$\times 10^{-2}$ &  &
{\small 9.14}$_{-0.97}^{+1.15}\times 10^{27}$ &  & {\small 1.05} \\
$^{96}${\small Zr} &  & {\small 2.603}$\times 10^{-15}$ &  & {\small 1.9}$%
\times 10^{21}$ & \cite{argy10}~~~~ & {\small (a)} &  & {\small 1.49}$\times
10^{-4}$ &  & {\small 4.19}$_{-0.30}^{+0.34}\times 10^{25}$ &  & {\small %
15.45} \\
&  &  &  &  &  & {\small (b)} &  & {\small 2.08}$\times 10^{-4}$ &  &
{\small 8.20}$_{-0.62}^{+0.70}\times 10^{25}$ &  & {\small 11.04} \\
$^{98}${\small Mo} &  & {\small 2.926}$\times 10^{-21}$ &  &  &  & {\small %
(a)} &  &  &  & {\small 7.36}$_{-0.64}^{+0.74}\times 10^{30}$ &  & {\small %
0.04} \\
&  &  &  &  &  & {\small (b)} &  &  &  & {\small 1.44}$_{-0.14}^{+0.16}%
\times 10^{31}$ &  & {\small 0.03} \\
$^{100}${\small Mo} &  & {\small 1.736}$\times 10^{-15}$ &  & {\small 2.7}$%
\times 10^{22}$ & \cite{arno11}~~~~ & {\small (a)} &  & {\small 2.22}$\times
10^{-5}$ &  & {\small 1.33}$_{-0.16}^{+0.20}\times 10^{25}$ &  & {\small %
27.45} \\
&  &  &  &  &  & {\small (b)} &  & {\small 3.09}$\times 10^{-5}$ &  &
{\small 2.58}$_{-0.33}^{+0.40}\times 10^{25}$ &  & {\small 19.69} \\
$^{104}${\small Ru} &  & {\small 3.037}$\times 10^{-17}$ &  &  &  & {\small %
(a)} &  &  &  & {\small 1.55}$_{-0.18}^{+0.21}\times 10^{27}$ &  & {\small %
2.54} \\
&  &  &  &  &  & {\small (b)} &  &  &  & {\small 2.97}$_{-0.36}^{+0.44}%
\times 10^{27}$ &  & {\small 1.83} \\
$^{110}${\small Pd} &  & {\small 2.758}$\times 10^{-16}$ &  &  &  & {\small %
(a)} &  &  &  & {\small 6.39}$_{-0.83}^{+1.03}\times 10^{25}$ &  & {\small %
12.51} \\
&  &  &  &  &  & {\small (b)} &  &  &  & {\small 1.25}$_{-0.17}^{+0.21}%
\times 10^{26}$ &  & {\small 8.95} \\
$^{128}${\small Te} &  & {\small 9.669}$\times 10^{-18}$ &  & {\small 1.5}$%
\times 10^{24}$ & \cite{manu91,bar10b}~~~~ & {\small (a)} &  & {\small 6.88}$%
\times 10^{-5}$ &  & {\small 7.09}$_{-0.93}^{+0.12}\times 10^{27}$ &  &
{\small 1.19} \\
&  &  &  &  &  & {\small (b)} &  & {\small 9.54}$\times 10^{-5}$ &  &
{\small 1.36}$_{-0.18}^{+0.22}\times 10^{28}$ &  & {\small 0.86} \\
$^{130}${\small Te} &  & {\small 1.300}$\times 10^{-15}$ &  &  &  & {\small %
(a)} &  &  &  & {\small 4.23}$_{-0.68}^{+0.90}\times 10^{25}$ &  & {\small %
15.37} \\
&  &  &  &  &  & {\small (b)} &  &  &  & {\small 8.18}$_{-0.13}^{+0.18}%
\times 10^{25}$ &  & {\small 11.05} \\
$^{150}${\small Nd} &  & {\small 1.038}$\times 10^{-14}$ &  & {\small 1.52}$%
\times 10^{21}$ & \cite{argy09}~~~~ & {\small (a)} &  & {\small 8.50}$\times 10^{-5}$ &  &
{\small 1.10}$_{-0.24}^{+0.36}\times 10^{25}$ &  & {\small 30.19} \\
&  &  &  &  &  & {\small (b)} &  & {\small 1.18}$\times 10^{-4}$ &  &
{\small 2.12}$_{-0.47}^{+0.71}\times 10^{25}$ &  & {\small 21.73} \\
\hline\hline
\end{tabular}
\end{table*}

\subsection{Sterile neutrinos}

To estimate the uncertainties in NTMEs $M^{\left( 0\nu \right) }(m_{h})$,
sets of twenty-four NTMEs are calculated for five potential $\left( \beta
^{-}\beta ^{-}\right) _{0\nu }$ candidates, namely $^{96}$Zr, $^{100}$Mo, $%
^{128,130}$Te and $^{150}$Nd isotopes using HFB wavefunctions generated with
four different parametrizaions of effective two-body interaction, form factors 
with two different parametrizations and three
different parameterizations of the Jastrow type of SRC in the mass range $%
m_{h}=10^{-4}$ MeV--$10^{9}$ MeV. It is noticed that in the limit $%
m_{h}\rightarrow 0$, $M^{\left( 0\nu \right) }(m_{h})\rightarrow M^{\left(
0\nu \right) }$ and $M^{\left( 0\nu \right) }(m_{h})\rightarrow \left(
m_{p}m_{e}/m_{h}^{2}\right) M^{\left( 0N\right) }$ in the limit $%
m_{h}\rightarrow $l$\arg $e. The extracted averages $\overline{M}^{(0\nu
)}(m_{h})$ and uncertainties $\Delta \overline{M}^{(0\nu )}(m_{h})$ are
presented in Table~\ref{tab8}. The uncertainties  
$\Delta \overline{M}^{(0\nu )}(m_{h})$ vary in between 4\% (9\%)--20\% (36\%) 
without (with) SRC1 depending on the considered mass of the sterile neutrinos. 
The extracted limits on the $\nu _{h}-\nu _{e}$
mixing matrix element $U_{eh}$ from the largest observed limits on the
half-lives $T_{1/2}^{0\nu }$ of $\left( \beta ^{-}\beta ^{-}\right) _{0\nu }$
decay are displayed in Fig. 3. The extracted limits on the $\nu _{h}-\nu _{e}$
mixing matrix element $U_{eh}$ span a wider region of $\nu _{h}$ mass $m _{h}$
than those of laboratory experiments, astrophysical and cosmological 
observations \cite{dogo00} and are comparable to the limits obtained in Ref. 
\cite{blen10}.

\begin{figure}[htbp]
\includegraphics [scale=0.30]{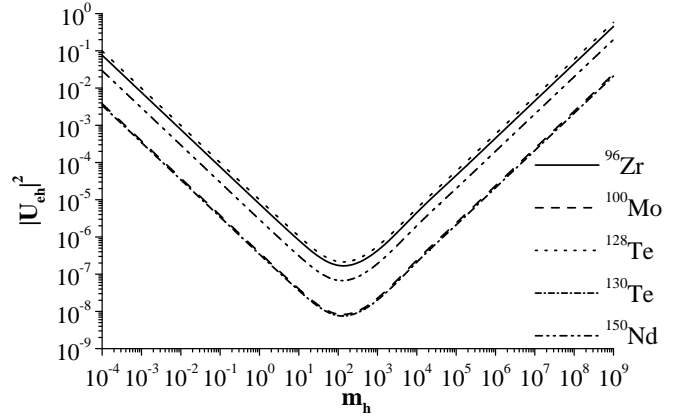}
\caption{Variation in extracted limits on the $\nu _{h}-\nu _{e}$
mixing matrix element $\left|U_{eh}\right|^2$ with the mass $m_{h}$.}
\end{figure}

\begin{table*}[htbp]
\caption{Calculated average NTMEs $\Delta \overline{M}^{(0\nu )}(m_{h})$ and uncertainties 
 $\Delta \overline{M}^{(0\nu )}(m_{h})$ for the 
$\left( \beta ^{-}\beta ^{-} \right) _{0\nu }$ decay
of $^{94,96}$Zr, $^{98,100}$Mo, $^{104}$Ru, $^{110}$Pd, $^{128,130}$Te and $%
^{150}$Nd isotopes. The (a) Case I and (b) Case II denote calculations with and without 
SRC1, respectively.}
\label{tab8}
\begin{tabular}{cccccccccccccccc}
\hline\hline
{\small Nuclei} &~~~~~~~~~~~  & $
\begin{array}{c}
m_{h}=~~~ \\
(MeV)
\end{array}
$ & {\small 10}$^{-4}-1$ &  & ~~~~~~~ & {\small 10} &  & ~~~~~~~ & {\small 10}$^{2}$ &  &
~~~~~~~& {\small 10}$^{3}$ &  & ~~~~~~~ & $
\begin{array}{c}
10^{4}-10^{9} \\
\times 10^{5}/m_{h}^{2}
\end{array}
$ \\ \hline
$^{96}${\small Zr} & {\small (a)} &  & {\small 2.863}$\pm ${\small 0.260} &
&  & {\small 2.692}$\pm ${\small 0.252} &  &  & {\small 1.273}$\pm ${\small %
0.170} &  &  & {\small 0.050}$\pm ${\small 0.015} &  &  & {\small 0.486}$\pm
 ${\small 0.175} \\
& {\small (b)} &  & {\small 3.028}$\pm ${\small 0.116} &  &  & {\small 2.853}%
$\pm ${\small 0.109} &  &  & {\small 1.384}$\pm ${\small 0.067} &  &  &
{\small 0.060}$\pm ${\small 0.008} &  &  & {\small 0.593}$\pm ${\small %
0.102} \\
$^{100}${\small Mo} & {\small (a)} &  & {\small 6.262}$\pm ${\small 0.626}
&  &  & {\small 5.910}$\pm ${\small 0.597} &  &  & {\small 2.808}$\pm $%
{\small 0.360} &  &  & {\small 0.105}$\pm ${\small 0.030} &  &  & {\small %
0.998}$\pm ${\small 0.347} \\
& {\small (b)} &  & {\small 6.589}$\pm ${\small 0.438} &  &  & {\small 6.229}%
$\pm ${\small 0.405} &  &  & {\small 3.028}$\pm ${\small 0.188} &  &  &
{\small 0.124}$\pm ${\small 0.016} &  &  & {\small 1.210}$\pm ${\small 0.204}
\\
$^{128}${\small Te} & {\small (a)} &  & {\small 3.619}$\pm ${\small 0.388} &
&  & {\small 3.412}$\pm ${\small 0.374} &  &  & {\small 1.640}$\pm ${\small %
0.239} &  &  & {\small 0.064}$\pm ${\small 0.019} &  &  & {\small 0.613}$\pm
 ${\small 0.220} \\
& {\small (b)} &  & {\small 3.819}$\pm ${\small 0.278} &  &  & {\small 3.607}%
$\pm ${\small 0.263} &  &  & {\small 1.776}$\pm ${\small 0.153} &  &  &
{\small 0.076}$\pm ${\small 0.011} &  &  & {\small 0.744}$\pm ${\small 0.138}
\\
$^{130}${\small Te} & {\small (a)} &  & {\small 4.054}$\pm ${\small 0.488} &
&  & {\small 3.815}$\pm ${\small 0.450} &  &  & {\small 1.808}$\pm ${\small %
0.225} &  &  & {\small 0.069}$\pm ${\small 0.019} &  &  & {\small 0.659}$\pm
 ${\small 0.223} \\
& {\small (b)} &  & {\small 4.262}$\pm ${\small 0.392} &  &  & {\small 4.018}%
$\pm ${\small 0.349} &  &  & {\small 1.949}$\pm ${\small 0.104} &  &  &
{\small 0.081}$\pm ${\small 0.010} &  &  & {\small 0.796}$\pm ${\small 0.127}
\\
$^{150}${\small Nd} & {\small (a)} &  & {\small 2.831}$\pm ${\small 0.422} &
&  & {\small 2.659}$\pm ${\small 0.396} &  &  & {\small 1.213}$\pm ${\small %
0.198} &  &  & {\small 0.043}$\pm ${\small \ 0.013} &  &  & {\small 0.413}$%
\pm ${\small 0.149} \\
& {\small (b)} &  & {\small 2.963}$\pm ${\small 0.395} &  &  & {\small 2.788}%
$\pm ${\small 0.368} &  &  & {\small 1.303}$\pm ${\small 0.161} &  &  &
{\small 0.051}$\pm ${\small 0.008} &  &  & {\small 0.500}$\pm ${\small 0.098}
\\ \hline\hline
\end{tabular}
\end{table*}

\section{CONCLUSIONS}

In the PHFB model, the required NTMEs to study the 
$\left( \beta ^{-}\beta ^{-}\right) _{0\nu }$ \ decay of $^{94,96}$Zr, 
$^{98,100}$Mo, $^{104}$Ru, $^{110}$Pd, $^{128,130}$Te and $^{150}$Nd
isotopes within mechanisms involving the light Majorana neutrino, classical 
Majorons and sterile neutrinos  are calculated for four different
parametrization of pairing plus multipolar type of effective two body
interaction, two sets of form factors and three different parametrizations of SRC. 
It is observed that the closure approximation is quite valid as expected. 
The effect due to FNS is about 10\% and inclusion of the HOC further reduces the 
NTMEs by approximately 12\%. With the consideration of the SRCs, the NTMEs are 
in addition reduced by 16.0\% (1.0\%) for SRC1 (SRC2). The effects due to deformation 
are in between a factor of 2--6.

The sets of twenty-four NTMEs $M^{\left( 0\nu \right) }$ have been employed
for estimating the uncertainties therein for the bare axial vector coupling
constant $g_{A}=1.254$ and quenched value of $g_{A}=1.0$. In the mechanisms involving 
light Majorana neutrino and classical Majorons, the uncertainties in NTMEs are in 
about 4.0\% (9.0\%)--13.5\% (15.0\%) without (with) the SRC1. In the case of sterile neutrinos, 
the uncertainties in NTMEs are in between 4\% (9\%)-- 20\% (36\%) depending on the 
considered mass of the sterile neutrinos without (with) SRC1. 

We have also extracted limits on the effective neutrino mass 
$\left\langle m_{\nu }\right\rangle $ from the
available limits on experimental half-lives $T_{1/2}^{0\nu }$ using average
NTMEs $\overline{M}^{(0\nu )}$ calculated in the PHFB model. In the case of $%
^{130}$Te isotope, one obtains the best limit on the effective neutrino mass 
$\left\langle m_{\nu }\right\rangle$ $ < $ 0.32 eV--0.45 eV from the observed limit on 
the half-live $T_{1/2}^{0\nu }$ $ > $ $3.0 \times 10^{24}$ yr of $%
\left( \beta ^{-}\beta ^{-}\right) _{0\nu }$ decay \cite{arna08}. The best limit on 
the Majoron-neutino coupling constant $\left\langle g_{M }\right\rangle$ turns out 
to be $< $ 2.22 $\times 10 ^{-5}$ from the observed half-life of $^{100}$Mo isotope. 
The study of sensitivities of nuclei suggest that to extract the effective mass of 
light Majorana neutrino $\left\langle m_{\nu }\right\rangle$,  $^{100}$Mo is the 
preferred isotope and  $^{150}$Nd is the favorable isotope to extract the 
$\left\langle g_{M }\right\rangle$. Thus, the sensitivities of different nuclei 
are mode dependent. It has been observed that the extracted limits on the sterile 
neutrino $\nu _{h}-\nu _{e}$ mixing matrix element $U_{eh}$ extend over a wider region of 
mass $m _{h}$ than those of laboratory experiments, astrophysical and cosmological 
observations.

\begin{acknowledgments}
This work is partially supported by the Council of Scientific and Industrial
Research (CSIR), India vide sanction No. 03(1216)/12/EMR-II, Indo-Italian
Collaboration DST-MAE project via grant no. INT/Italy/P-7/2012 (ER), Consejo
Nacional de Ciencia y Tecnolog\'{i}a (Conacyt)-M\'{e}xico,
and Direcci\'{o}n General de Asuntos del Personal Acad\'{e}mico,
Universidad Nacional Aut\'{o}noma de M\'{e}xico (DGAPA-UNAM) project IN103212. 
One of us (PKR) thanks Prof. S. K. Singh, HNBG University, India for useful 
discussions.
\end{acknowledgments}

\end{document}